\documentclass[reprint,amsmath,amssymb, aps, pra,superscriptaddress]{revtex4-1}

\usepackage{graphicx}
\usepackage{epsfig}
\usepackage{color}
\usepackage{amsmath}
\usepackage{amsfonts}
\usepackage{mathrsfs}
\usepackage{txfonts}
\usepackage{color}

\usepackage[breaklinks]{hyperref} 
\hypersetup{colorlinks=true, linkcolor=blue, citecolor=blue, filecolor=blue, urlcolor=blue}

\begin{document}
	\title{Moire Localization in Two Dimensional Quasi-Periodic Systems}
	\author{Biao Huang}
	\email{phys.huang.biao@gmail.com} 
	\affiliation{Department of Physics and Astronomy, University of Pittsburgh, Pittsburgh PA 15260, USA}
	\author{W. Vincent Liu}
	\email{wvliu@pitt.edu}
	\affiliation{Department of Physics and Astronomy, University of Pittsburgh, Pittsburgh PA 15260, USA}
	\affiliation{
		Wilczek Quantum Center, School of Physics and Astronomy and T. D. Lee Institute,  Shanghai Jiao Tong University, Shanghai 200240, China}
	\date{\today}
	
	\begin{abstract}
		We discuss a two-dimensional system under the perturbation of a Moire potential, which takes the same geometry and lattice constant as the underlying lattices but mismatches up to relative rotation. Such a self-dual model belongs to the orthogonal class of a quasi-periodic system whose features have been evasive in previous studies. We find that such systems enjoy the same scaling exponent as the one-dimensional Aubry-Andre model $ \nu\approx 1 $, which saturates the Harris bound $ \nu>2/d=1 $ in two-dimensions. Meanwhile, there exist a continuous and rapid change for the inverse participation ratio in the eigenstate-disorder plane, different from typical one-dimensional situation where only a few or no step-like contours show up. An experimental scheme based on optical lattices is discussed. It allows for using lasers of arbitrary wavelengths and therefore is more applicable than the one-dimensional situations requiring laser wavelengths close to certain incommensurate ratios.
	\end{abstract}
	\maketitle
	
	\section{Introduction}
	
	Overlaying lattices with mismatches leads to the so-called Moire pattern, which has long been an intriguing subject in two-dimensional heterostructures~\cite{Neto2009,Dean2013,Hunt2013}. Usually, a commensurate twist angle between two layers are considered, such that the bilayer system still possesses an enlarged periodicity. The interlayer hopping and interactions would then induce a reconstruction of the original Bloch waves in each layer into multiple bands. That provides a new knob to tune the property of systems with various twisting angles, leading to novel examples including the flat-band induced superconductivity for bilayer graphene twisted at magic angle~\cite{Cao2018,Yankowitz_2019}. More recently, the experimental advancement has made it possible to stabilize the bilayer graphene system at incommensurate large twist angle~\cite{Yao2018}. In such situations, the combined bilayer system breaks any translation symmetry and the Bloch waves in each layer are destroyed completely. Then, it is of interest to ask what may be the general phenomena expected in these circumstances. 
	
	The incommensurately twisted bilayer system resembles quasi-crystals, in the sense that only rotation but not translation symmetry is preserved therein. In the case of quasi-crystals, localization of particles has been a focus of study for decades~\cite{Sokoloff1986,Krajci1995,Villa2006,Park2018}. One could relate the two systems using the following schematic reasoning. Consider a bilayer, two-dimensional system with interlayer interactions,
	\begin{align}\label{eq:fullh}
	H_{\text{bi}} = \sum_{\langle i,j\rangle} (a_i^\dagger a_j + b_i^\dagger b_j) + \sum_{i,j}V_{ij} n_i^{A} n_j^{B}, 
	\end{align}
	where $ a_i^\dagger, b_i^\dagger $ create particles in two layers respectively, and $ n_i^{A} = a_i^\dagger a_i, n_i^{B} = b_i^\dagger b_i $.
	Now, suppose one could take as a starting point the decomposition $ \mu_i^A = \sum_j V_{ij} \langle n_j^B\rangle $, and $ \mu_j^B = \sum_i V_{ij}\langle n_i^A \rangle $, then the above model reduces to
	\begin{align}\label{eq:heff}
	H_{\text{eff}} = \sum_{\langle i,j\rangle} a_i^\dagger a_j + \sum_i \mu_i^A n_i^A + (A\leftrightarrow B),
	\end{align}
	which describes a decoupled bilayer under onsite chemical potentials $ \mu_i^{A,B} $ respectively. With reasonable interactions, i.e. not infinite-range interaction with identical strength, one would expect an incommensurate twist angle to result in an incommensurate, quasi-periodic pattern for $ \mu_i^{A,B} $. That gives an analogous scenario as in quasi-crystals, with the difference that here, it is the chemical potentials rather than lattice site locations that break translation symmetry. 
	
	The inter-sample interaction induced slow-relaxation (``quasi many-body localization") has previously been demonstrated for a one-dimensional translation invariant ladder by Yao {\em et. al.}~\cite{Yao2016}. Specifically, given a highly non-uniform initial density distribution and strong interactions, the particles in one chain would serve as random chemical potential for the other, similar to the case mentioned above. But in contrast, due to the incommensurate twist angles breaking translation symmetry here, it is expected a genuine localization without initial state dependence would occur for incommensurate Moire bilayers. In particular, even if the particles are uniformly distributed in one layer, they could still function as disorder potentials for the other through quasi-periodic $ V_{ij} $. 
	
	Thus, we are motivated to investigate the ``Moire localization" possibly held in a two-dimensional quasi-periodic system. As a first step, we focus on an effective scenario in Eq.~(\ref{eq:heff}), namely, a regular single-layer lattice under the perturbation of quasi-periodic potentials. This would serve as a good starting point for further taking into account fluctuations in Eq.~(\ref{eq:fullh}). Also, although the localization of quasi-periodic systems in one-dimension have been thoroughly studied for decades~\cite{Nandkishore2015,Abanin2018,Khemani2017,Roati2008,Schreiber2015,Choi2016}, since the renowned work by Aubry and Andre~\cite{Aubry1980} in 1980, its generalization to higher dimensions has just started quite recently~\cite{Bordia2017a,Devakul2017,Sutradhar2018}. The theoretical studies on this regard have remained chiefly in the single-particle level, with many crucial aspects still open to discussions. In particular, the scaling exponents for {\em orthogonal} class of models in {\em two-dimensions} have been left out in Ref.~\cite{Devakul2017,Sutradhar2018} due to various difficulties. But this class of models are most relevant for experiments on genuine two-dimensional quasi-periodic system with real hopping, and also for the theoretical analysis based on Eqs.~(\ref{eq:fullh}) and (\ref{eq:heff}). Therefore, it is useful to clarify the relevant single-particle physics before considering the full many-body interactions.
	
	Apart from preparing for analysis of interaction-induced localization, generalizing the paradigmatic Aubry-Andre model to higher dimensions itself may yield interesting results. As shown by Devakul and Huse in Ref.~\cite{Devakul2017}, for orthogonal class of models in three dimensions, the scaling exponents for (single-particle) localization transitions are the same for both quasi-periodic and purely random potentials. This is in contrast to the situation in one-dimension, where the quasi-periodic Aubry-Andre model gives rise to the scaling exponent strongly violating the Harris bound~\cite{Harris1974} formulated for purely random models. Such a violation in one-dimension is indicated in Ref.~\cite{Khemani2017} to persist into the many-body localization scenario.  Experimentally, a higher-dimensional quasi-periodic potential is actually more suitable for cold atom experiment. This is because the property of the system is most sensitive to the relative rotation angle between the Moire perturbing potential and the main lattice potential, rather than the relative lattice constants. That means one could use lasers of any frequency available in the laboratory to engineer the desirable system. Finally, having a good understanding of the single particle physics for quasi-periodic systems would pave the way for possible perturbative treatment~\cite{Basko2006} of many-body localization therein in the future.
	
	In this work, we study a two-dimensional square lattice under the perturbation of Moire-type of potential, which takes exactly the same lattice geometry and constants as the underlying lattice. A commensuration condition for the relative rotation angles between the two lattice potentials is provided. Such a condition also helps identifying ``better" rotation angles that are further from a commensurate one, resembling the ``better" ratio of lattice constants in a one-dimensional quasi-periodic system, that is usually taken to be the golden ratio. Two important features of such a system is revealed. First, for the inverse-participation-ratios in the eigenstate-disorder plane, in contrast to the one-dimensional situation where only a few step-like jumps exist, our two-dimensional model exhibits a rapid and continuous change. Such a character indicates the unusual mobility edges in two dimensions. Second, the critical exponent for localization length is extracted from the multifractal analysis, with the value $ \nu\approx1 $ saturating the Harris bound $ \nu>2/d=1 $ for two-dimensions. Finally, we provide an experimental scheme to realize such a Moire type of model, where lasers of arbitrary wavelengths could be applied.

	\section{Models and the Commensuration Conditions}\label{sec:comm}
	
	Consider a square lattice under perturbations of Moire-type of potentials with the same geometry as the underlying major one. There are three types of controlling parameters for such a perturbing potential. (a) Stretching: it could possess different lattice constants. (b) Rotation: it may be rotated by an angle $ \theta $ with respect to the underlying lattice. (c) Translation: there may be a global relative translation between these two lattices. Both (a) and (b) could generate an incommensurate potential, but as in the experiment Ref.~\cite{Bordia2017a}, without interactions, (a) alone would result in a separable $ V(x,y) = V(x) + V(y) $ such that the model is reduced to two orthogonal lower-dimensional ones. Meanwhile, (c) does not change the universal properties of the system in thermodynamic limits, but only affects microscopic details for a finite size system, i.e. the wave functions in the boundary. Thus, the global rotation angle of perturbing potential is the most important factor characterizing a genuine two-dimensional quasi-periodic system. It was also found in previous works that the system's properties are most sensitive to rotation angles of the perturbing potentials~\cite{Devakul2017,Sutradhar2018}.
	
	Thus, we focus on the following model with two overlapping square lattices, where the stronger one gives rise to the tight-binding approximation, and the weaker one generates the Moire type of onsite chemical potentials,
	\begin{align}\nonumber
	H &= -\sum_{\langle i,j\rangle} (c_i^\dagger c_j+ c_j^\dagger c_i) + V_d \sum_i \mu_i c_i^\dagger c_i,
	\\ \nonumber
	\mu_i &= 
	\sin^2 \left[\pi u_i - \varphi_1 \right] + \sin^2 \left[\pi v_i - \varphi_2 \right],\\\nonumber
	 u_i &= x_i\cos\theta  -y_i\sin\theta,\qquad 
	v_i = x_i\sin\theta  + y_i\cos\theta,
	\\\label{moire}
	\varphi_1 &= \pi (a\cos\theta  -b\sin\theta),\qquad 
	\varphi_2 = \pi(a\sin\theta  + b\cos\theta).
	\end{align}
	Here $ i=(x_i, y_i) $ denotes the square lattice sites, with $ x_i,y_i \in \mathbb{Z} $. The Moire superlattice potential $ \mu_i $, compared with the main lattice, is rotated by an angle $ \theta $ at the origin $ (x,y)=(0,0) $ and then translated by $ (x,y) = (a,b) $. The stretching (a) would be neglected in most parts below, because it does not lead to extra features and only complicates the analysis. For instance, either stretching the lattice constant by a factor $ \sqrt2 $, or a rotation of $ \pi/4 $ would produce an incommensurate potential, while combining these two only gives a usual commensurate, staggered lattice. We would chiefly investigate the influence of different rotation angles $ \theta $, and average over various translations $ (a,b) $ (or equivalently, $ \varphi_1, \varphi_2 $) when extracting universal properties from finite size numerical results.

	\begin{figure}
		[h]
		\parbox{3.6cm}{
			\includegraphics[width=3.5cm]{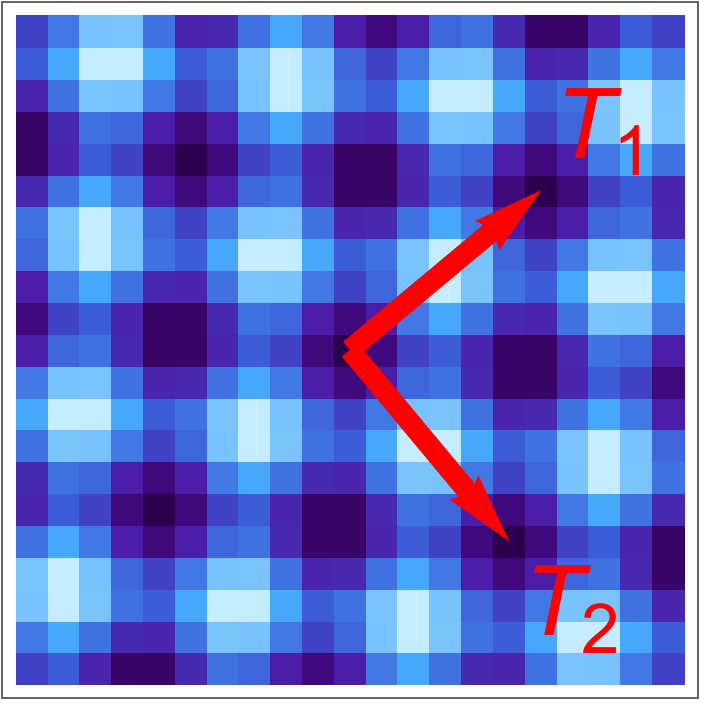} \\ (a) Commensurate }
		\parbox{3.6cm}{
			\includegraphics[width=3.5cm]{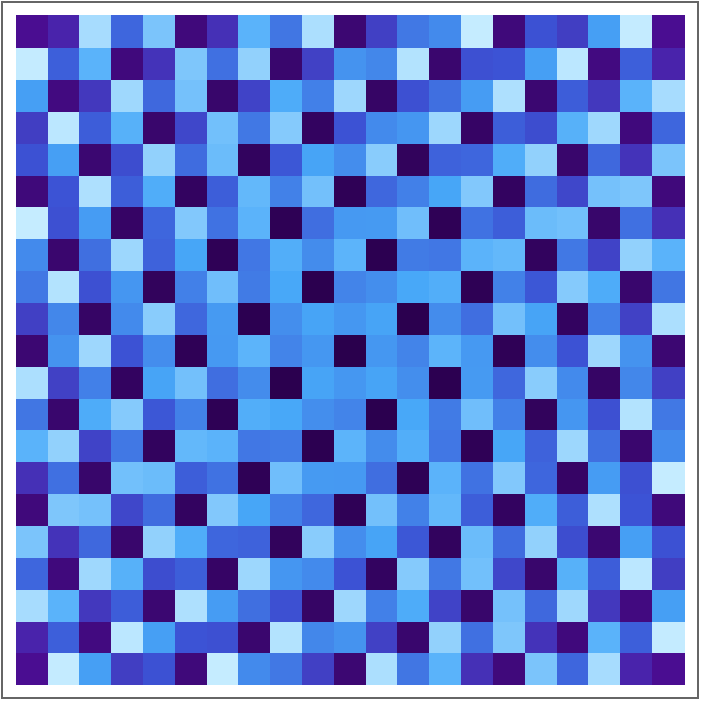} \\ (b) Incommensurate}
		\parbox{0.6cm}{
			\includegraphics[width=0.8cm]{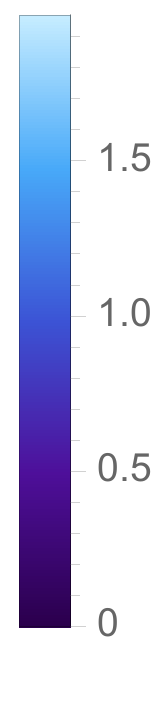}}
		\caption{\label{fig:moire} The Moire potential pattern $ \mu_i $ for (a) the commensurate angle $ \theta=\arccos(60/61) $ ($ q=11, p=1 $), and (b) the incommensurate angle $ \theta=\pi/5 $. Each pixel denotes one lattice site, and the color indicates the strength of the disorder potential therein.}
	\end{figure}
	The first question to ask is under what condition would the rotation angle $ \theta $ result in a commensurate perturbing potential $ \mu_i $. Similar to the analysis in honeycomb lattices~\cite{Shallcross2010,Santos2012,Yao2018}, the commensuration condition is prescribed by solving a Diophantine equation specified by lattice geometries. For square lattices, the solution is readily provided by the Pontryagin's triples~\cite{Steuding2005} (see Appendix \ref{app:comm} for derivations)
	\begin{align}\label{commCond}
	\theta_{\text{c}} = \frac{m\pi}{2} \pm \arccos\frac{q^2-p^2}{q^2+p^2},\qquad m\in Z
	\end{align}
	where $ q, p $ are positive, coprime integers. This is the commensurate condition for Moire rotation angles of rectangular lattices with rational aspect ratios $ \gamma $. Specific to the square lattice $ \gamma=1 $, due to four-fold rotation symmetry and mirror symmetry along $ \hat{x}\pm\hat{y} $ directions, one only needs to consider $ \theta_c = \arccos\frac{q^2-p^2}{q^2+p^2}\in(0,\pi/4) $ and $ \theta_c'=\pi/2-\theta_c = \arccos\frac{2qp}{q^2+p^2} \in(0,\pi/4) $, with coprime $ q>p $. The resulting Moire pattern has periodic structures, with the new Moire-Bravais lattice vectors $ \boldsymbol{T}_1, \boldsymbol{T}_2 $ given below. When $ p $ or $ q $ is an even number (they cannot both be even due to the coprime requirement),
	\begin{align}\label{t1t2}
	\boldsymbol{T}_1 = q\hat{x} - p\hat{y}, &&
	\boldsymbol{T}_2 = p\hat{x} + q\hat{y},
	\end{align}
	and when $ (q,p) $ are both odd numbers,
	\begin{align}\label{t1t2odd}
	\boldsymbol{T}_1 = \frac{q+p}{2}\hat{x} + \frac{q-p}{2}\hat{y} , &&
	\boldsymbol{T}_2 = \frac{q-p}{2} \hat{x} - \frac{q+p}{2} \hat{y}.
	\end{align}
	In both cases, the two vectors have identical length $ |\boldsymbol{T}_1|=|\boldsymbol{T}_2|=\sqrt{q^2+p^2} $ (when $ q $ or $ p $ is even) or $ \sqrt{(q^2+p^2)/2} $ (when $ q, p $ are both odd), and are orthogonal to each other $ \boldsymbol{T}_1 \cdot \boldsymbol{T}_2 = 0 $.
	An example of commensurate Moire pattern is drawn in Fig.~\ref{fig:moire}(a), where $ q=11, p=1 $, and the Bravais vectors are $ \boldsymbol{T}_1 = 6\hat{x} + 5\hat{y}$, $\boldsymbol{T}_2= 5\hat{x} - 6\hat{y} $. In contrast, an incommensurate example is shown in Fig.~\ref{fig:moire}(b) where only the four-fold rotation, but not any translation symmetry, exists.
	
	Here, we are interested in localization induced by incommensurate potentials, so it is worthwhile to choose among different incommensurate angles for further numerical analysis. Strictly speaking, any angle other than those given by Eq.~(\ref{commCond}), i.e. any $ \theta=\arccos(A) $ with $ A $ (or $ A $ module $ \pi/2 $) being irrational numbers, would render incommensurate systems belonging to the same universal class. But when one is trying to extract information from numerical results in a finite-size system, there is a subtle question of ``how incommensurate" the potentials are. Mathematically, it means if one takes a series of commensurate angles $ \{\theta_n, n=1,2,3\dots\} $, whose $ n\rightarrow\infty $ gives an incommensurate angle $ \theta_{\infty} $, then a ``more incommensurate" angle would require each $ \cos\theta_n $ being given by larger $ (q,p) $ in Eq.~(\ref{commCond}). Equivalently put, since the length of Moire-Bravais vectors $ \sim\sqrt{q^2+p^2} $, that means given a system size (and therefore an upper limit of $ (q,p) $, since for larger $ (q,p) $, the system cannot cover even one Moire period), ``more incommensurate" $ \theta_\infty $'s are farther away from commensurate angles. This is in the same spirit as in one-dimension, where the ``golden-ratio" $ (\sqrt5-1)/2 $ --- the ``most irrational number" farthest away from rational ones --- is usually adopted in numerics as the relative ratios of lattice constants between main and perturbing lattices. 
	
	\begin{figure}
		[h]
		\includegraphics[width=5cm]{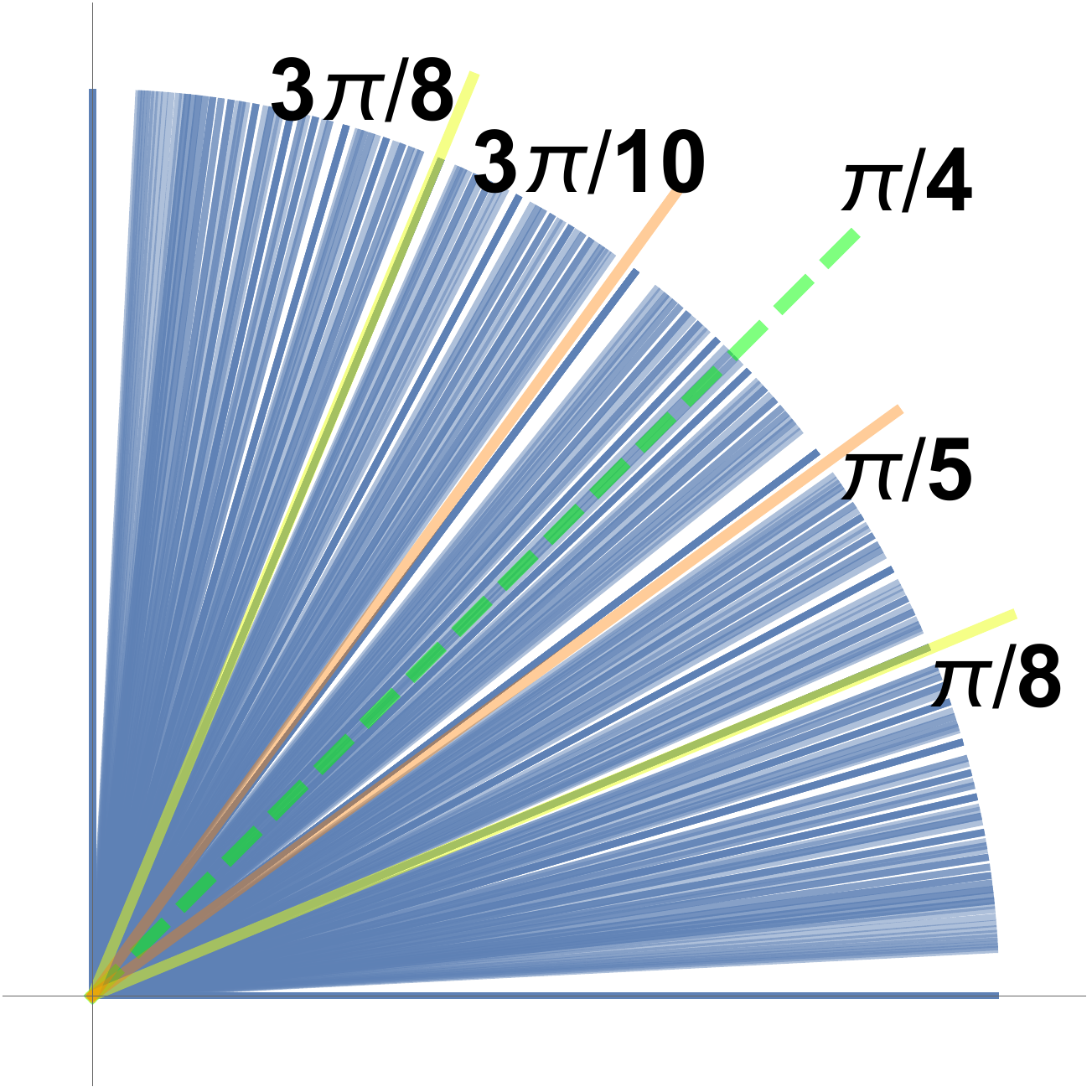}
		\caption{\label{fig:angles} The density of commensurate angles (blue polar lines), for $ 0 \le p \le q \le 20 $. Due to mirror symmetry along $ x=y $, the angles are symmetric along $ \theta=\pi/4 $. }
	\end{figure}
	To illustrate this point more clearly, we plot the commensurate angles for maximal $ (q,p)\le20 $ in Fig.~\ref{fig:angles}. We readily note that for all small angles $ \theta\rightarrow0 $ or certain large angles (i.e. $ \theta=\pi/5 $ or $ \theta=\pi/8 $), there appears less commensurate angles nearby. For small angles, there is a practical difference between bilayer heterostructure and optical lattice systems. For bilayer ones, there is a competition between interlayer attractive van der Waals force and elastic force. The former one tends to align the two layers by distorting lattices so as to minimize interlayer distance, while the latter one would favor keeping lattice shapes and therefore the interlayer twisting. For small twist angles $ \theta\rightarrow0 $, the van der Waals force overwhelms and the system could always be taken as a commensurate one~\cite{Woods2014}, with the enlarged periodicity approximately given by $ 1/\theta $. Although such an issue does not exist in optical lattices where lattice shapes are fixed, for generality of the results in practice, we consider the large twist angle $ \pi/5 $ for most parts in the following.
	
	It is worth emphasizing that even if one chooses an irrational lattice constant $ \beta $, there still exists the need to pick out twist angles further away from ``almost-commensurate" one so as to minimize finite size effects. The process there would consists of first finding a sequence of rational lattice constants close to $ \beta $, and then for each member in this sequence, find the density of commensurate angles. We would not digress to such a situation in this work.

	\section{Moire Localization}
	
	For the one-dimensional Aubry-Andre model~\cite{Aubry1980}, its self-duality gives rise to the unique localization transition (lack of mobility edge) and the uniform localization length. But if the incommensurate potential is given by relative twists rather than stretch, the two-dimensional eigenstates cannot be decomposed into orthogonal one-dimensional components. Then we would see that such a higher-dimensional generalization breaks the unique localization length and transition, in accordance with the three-dimensional results in Ref.~\cite{Devakul2017}. It can be traced back to that the Thouless formula adopted to prove these two points~\cite{Aubry1980}, which we review in Appendix \ref{app:AAmodel}, no longer holds for a higher-dimensional system. We would also illustrate the mobility edge clearly in numerics later on.
	
	\subsection{What does self duality imply in the Moire model?}\label{sec:duality}
	For later convenience, we rewrite Eq.~(\ref{moire}) as
	\begin{align}\nonumber
	H &= -\sum_{x,y} (c_{x+1,y}^\dagger +c_{x-1,y}^\dagger + c_{x,y+1}^\dagger + c_{x,y-1}^\dagger )c_{x,y} \\ \nonumber
	& \qquad - \frac{V_d}{2} \sum_{x,y} \mu_{x,y} c_{x,y}^\dagger c_{x,y},
	\\ \nonumber
	\mu_{x,y} &= 
	\cos \left[2\pi u_{x,y} + \varphi_1 \right] + \cos \left[2\pi v_{x,y} + \varphi_2 \right]\\ \nonumber
	u_{x,y} &= x\cos\theta  -y\sin\theta,\qquad 
	v_{x,y} = x\sin\theta  + y\cos\theta
	\\\label{newmoire}
	\varphi_1 &= 2\pi(a\cos\theta  -b\sin\theta) ,\qquad 
	\varphi_2 = 2\pi(a\sin\theta  + b\cos\theta)
	\end{align}
	which is equivalent to Eq.~(\ref{moire}) up to a constant. That also makes the model particle-hole symmetric after averaging over the phases $ \varphi_1,\varphi_2 $. The single-particle eigenfunction $ \phi_{x,y} = \langle x,y|\phi\rangle = \langle 0| c_{x,y}|\phi\rangle  $ satisfies 
	\begin{align}\label{real}
	-(\phi_{x+1,y} + \phi_{x-1,y} + \phi_{x,y+1} + \phi_{x,y-1}) - \frac{V_d}{2} \mu_{x,y} \phi_{x,y} = E \phi_{x,y},
	\end{align}
	where $ H|\phi\rangle = E|\phi\rangle $ and $ |0\rangle $ is vacuum.
	Consider the Fourier transformation
	\begin{align}\label{eq:fourier}
	\phi_{x,y} = e^{2\pi i (x\varphi_2 + y\varphi_1)} \sum_{m,n=-\infty}^\infty  \psi_{m,n} e^{2\pi i\left[ n(u_{x,y} + \varphi_1 ) + m(v_{x,y} + \varphi_2 ) \right] },
	\end{align}
	where $ m,n\in\mathbb{Z} $. Then, for incommensurate rotations where $ (m,n)\sin\theta$ or $ (m,n)\cos\theta $ are never integers and therefore different Fourier modes do not couple, the wave functions $ \psi_{m,n} $ satisfy the dual equation
	\begin{align}\nonumber
	&- 2\mu_{m,n}\psi_{m,n}  - \frac{V_d}{4} (\psi_{m+1,n} + \psi_{m-1,n} + \psi_{m,n+1} + \psi_{m,n-1} ) \\ \label{momentum}
	& = E \psi_{m,n}.
	\end{align}
	Compare Eqs.~(\ref{real}) and (\ref{momentum}), we see that the self-duality maps  $
	\frac{V_d}{2} \leftrightarrow \frac{8}{V_d},$ and $ E\leftrightarrow \frac{4E}{V_d}$.
	Then the invariant disorder strength under duality mapping can be extracted as 
	\begin{align}
	V_d^{(D)} = 4.
	\end{align}
	The dual Fourier transformation Eq.~(\ref{eq:fourier}) has the property that if $ |\psi_{m,n}|^2 $ is a localized wave function, in the sense that $ \sum_{m,n} |\psi_{m,n}|^2 $ and $ |\psi_{m,n}||_{m\text{ or }n\rightarrow\infty} $ is finite ($ |\psi_{m,n}| $ is exponentially localized around certain $ (m_0, n_0) $), we would have a delocalized $ \phi_{x,y} $ as $ \sum_{x,y}|\phi_{x,y}|^2\rightarrow\infty $~\footnote{
	Specifically, $ \sum_{x,y}|\phi_{x,y}|^2 = \sum_{x,y}\sum_{mnkl} \psi_{m,n}^* \psi_{k,l} A_{mnkl}^{x} B_{mnkl}^{y} C_{mnkl}  $, where $ A_{mnkl}^{x} = e^{2\pi i x[ (l-n)\cos\theta + (k-m)\sin\theta]} $, $ B_{mnkl}^{y} = e^{2\pi i y[-(l-n)\sin\theta + (k-m)\cos\theta]} $, and $ C_{mnkl} = e^{2\pi i [(l-n)\varphi_1 + (k-m)\varphi_2] } $. For $ m\ne k $ or $ n\ne l $, one could use $ \sum_{x=-L}^L e^{2\pi i \beta x} = \sin[\pi\beta(2L+1)]/\sin[\pi\beta] $ and show that $ D_{mnkl}\equiv C_{mnkl} \sum_{x,y}A_{mnkl}^xB_{mnkl}^y $ is finite as $ L\rightarrow\infty $. Then, $ \sum_{x,y}|\phi_{x,y}|^2 = \sum_{m\ne k \text{ or } n\ne l} \psi_{m,n}^* \psi_{k,l} D_{mnkl} + \sum_{mn} |\psi_{m,n}|^2 \sum_{x,y}1 \rightarrow \infty $, where the first term is finite due to the localization of $ \psi_{m,n} $.
	}. The boundary condition (or the normalization condition) for the wave function is that the maximal value of $ |\psi_{m_0,n_0}| $ at the localized site $ (m_0,n_0) $ is taken to be a constant. As such, the duality mapping implies a one-to-one correspondence between a localized eigenfunction at disorder $ V_d $ and energy $ E $, and a delocalized one at $ 16/V_d $ and energy $ 4E/V_d $. It does not specify which one is the localized solution.

	So far, the analysis is completely in parallel with that for the one-dimensional Aubry-Andre model. However, to prove all eigenstates are localized for $ V_d>V_d^{(D)} = 4 $ (and delocalized when $ V_d<V_d^{(D)} $), that is, a unique localization transition for all eigenstates, Ref.~\cite{Aubry1980} invoked the Thouless formula~\cite{Thouless1972}. Such a formula relates the localization length $ \xi_{E} $ at energy $ E $ with the density of states $ D(\varepsilon) $ as $ 1/\xi_E =  \int_{-\infty}^{\infty}  (\ln|E-\varepsilon|) D(\varepsilon) d \varepsilon $. Together with the duality mapping for density of states, one could show that all eigenstates are localized at the $ V_d>V_d^{(D)} $ side with uniform localization length $ 1/\xi = \ln(V_d/4) $ (see Appendix~\ref{app:AAmodel}). However, the Thouless formula, which requires the nearest-neighbor hopping in one-dimension (i.e. Eq.~(\ref{eq:EH}) and (\ref{eq:G1N})), no longer holds in higher dimensions. Then it is expected the mobility edge would generally exist for a higher dimensional model. 	
	
	In the following subsections, we perform exact diagonalization to further reveal the nature of the localization transitions in our Moire model.

	\subsection{Level statistics}
	
	We first compute the energy level statistics for the ratio of neighboring gaps~\cite{Oganesyan2007, Atas2013}. Arranging the energy levels $ \{E_m\} $ in the order $ E_m<E_{m+1} $, and defining the gap between neighboring levels in a finite-size system as $ \delta_m = E_{m+1} - E_m $, the ratio reads $ r_m = \min(\delta_{m},\delta_{m+1}) / \max(\delta_m, \delta_{m+1}) $. The distribution function of $ r_m $'s, $ P(r) $, would approach a Poisson limit $ P(r) = 2/(1+r)^2 $ with $ \langle r\rangle = \int_0^1 dr (rP(r)) \approx 0.386 $ for a fully localized system with complete sets of integrals of motion. In contrast, for delocalized systems, the neighboring energy level repulsions would lead to the Gaussian orthogonal ensemble (GOE) with $ \langle r\rangle \approx 0.5307 $. Ref.~\cite{Devakul2017} computed such a quantity for a fixed system size in three-dimensions, and found an ``intermediate'' regime near the dual-invariant disorder strength $ V_d^{(D)} $ with $ \langle r \rangle $ approaching the Gaussian ensembles. It is of interest to see whether such an intermediate phase, which would indicate the existence of mobility edge, persist to our two-dimensional situation. Also, we check the change of $ \langle r\rangle $ as the system size increases so as to indicate the behavior in the thermodynamic limit.
	
	The qualitative behavior of $ \langle r\rangle $ as a function of $ V_d $ can be expected as follows. At $ V_d=0 $, the translationally-invariant limit, momentums are conserved and $ E_{\boldsymbol{k}} = -2(\cos k_x + \cos k_y) $. Thus, it can be regarded as a localized system in momentum space. Deviating away from $ V_d = 0 $, the momentum conservation is immediately broken by the Moire perturbing potential, while real-space localization has not been established. Then, due to the lack of complete local integrals of motion, the level repulsion would lead to the GOE distribution for $ \langle r \rangle $. As $ V_d $ increases, more and more eigenstates are localized, and when the full localization for all eigenstates occur, $ \langle r \rangle $ would once again approach the Poisson limit with real-space positions as good quantum numbers to denote localized states.

	\begin{figure}
		[h]
		\includegraphics[width=8cm]{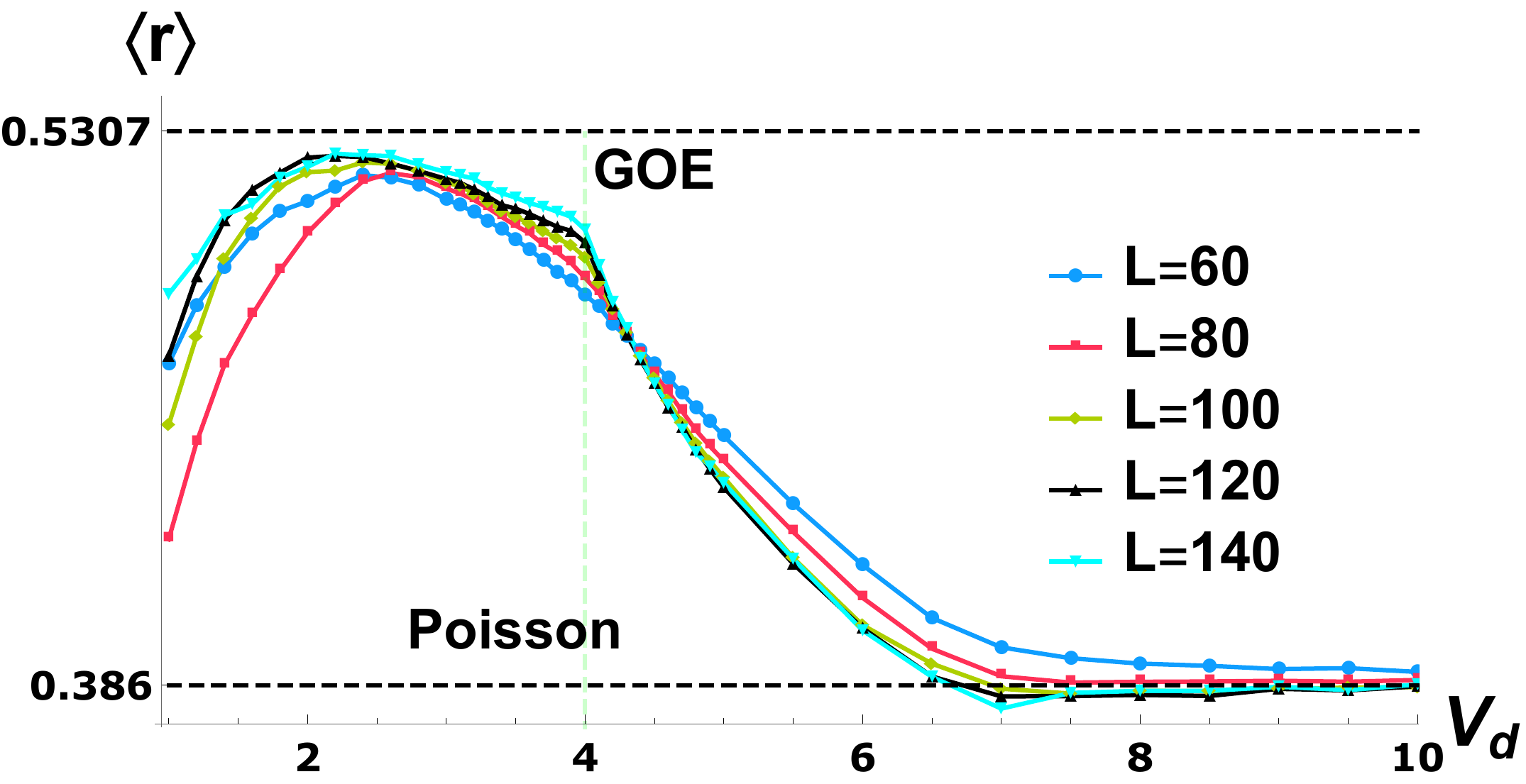}
		\caption{\label{fig:levelstat} The average value for neighboring level gap ratios. Here $ \theta=\pi/5 $. The numbers of phases $ \varphi_{1,2} $ averaged over are $ 4000/2000 $  $(L=60)$, $1000/500$ $(L=80)$, $200/100$ $(L=100)$, $100/50$ $(L=120)$, $60/20$ $(L=140) $ for $ V_d $ within/out-of the range $ [3,5] $. }
	\end{figure}
	
	The results for $ \theta=\pi/5 $ shown in Fig.~\ref{fig:levelstat} match the intuitive reasoning given above. Here, we avoid the subtle regime with small $ V_d \ll 1 $. This is because at $ V_d=0 $, the system possesses many symmetries, such as translation, reflection, and four-fold rotation symmetries. That renders the Hamiltonian block-diagonalized into different symmetry sectors and a meaningful level statistics requires the decomposition of Hilbert space into these sectors~\cite{Santos2010,Santos2010a,Sandvik2010,DAlessio2014}. Close to $ V_d=0 $, those symmetries are only weakly broken and a severe finite-size effect is expected. We readily see from Fig.~\ref{fig:levelstat} that at small $ V_d $, $ \langle r \rangle  $ decreases with irregular behaviors. On the other hand, the transition near $ V_d^{(D)} = 4 $ is well-behaved. At $ V_d=V_d^{(D)} = 4 $, $ \langle r\rangle $ keeps approaching the GOE limit as system sizes $ L $ increase. Passing $ V_d=4 $, $ \langle r\rangle $ starts to drop. We note that for relatively smaller system size (i.e. $ L=60, 80, 100 $), the curves cross at around $ V_d\approx4.4 $ with the increase of $ L $. However, for larger $ L $'s such a ``crossing'' appears to be replaced by a convergence of curves for different $ L $'s. We have also verified several other twisting angles and similar results emerge. Thus, the level statistics indicates the existence of mobility edge; only when $ V_d\gtrapprox 6.5 $ would almost all eigenstates be localized as $ \langle r\rangle $ approaches the Poisson limit.

	
	The average value $ \langle r \rangle $ gives an averaged behavior for all eigenstates. We next look at the eigenstate-resolved inverse participation ratio for more details regarding the mobility edge.

	\subsection{Inverse Participation Ratio and Mobility Edge}
	
\begin{figure}[h]
		\parbox{4cm}{
			\includegraphics[width=4cm]{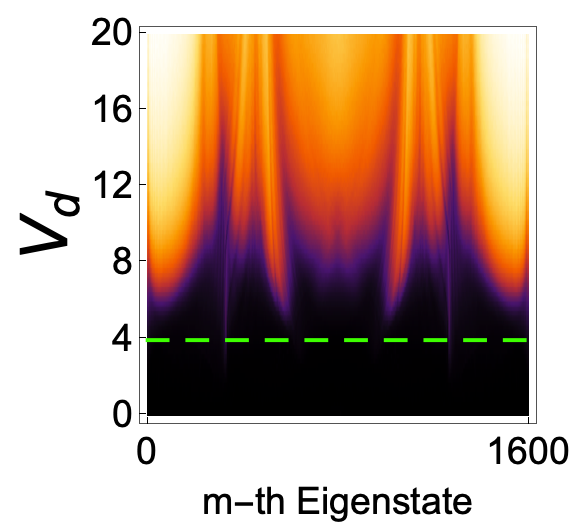}\\ (a) $ L=40 $  ($ 1000$ $ \varphi_{1,2} $'s) }
		\parbox{4cm}{
			\includegraphics[width=4cm]{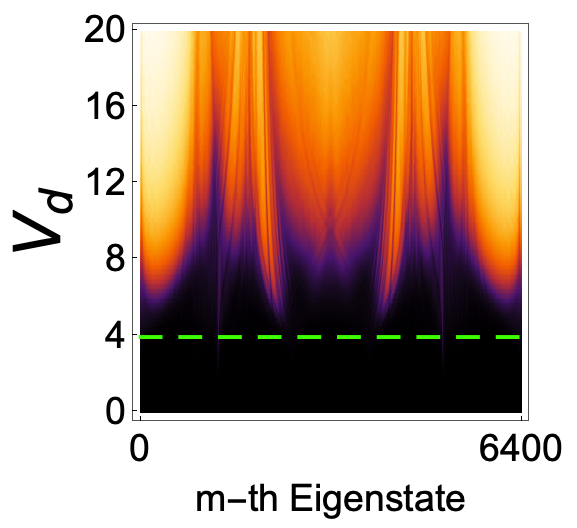}\\ (b) $ L=80 $ ($ 100$ $ \varphi_{1,2} $'s) }
		\parbox{0.5cm}{
		\includegraphics[width=0.4cm]{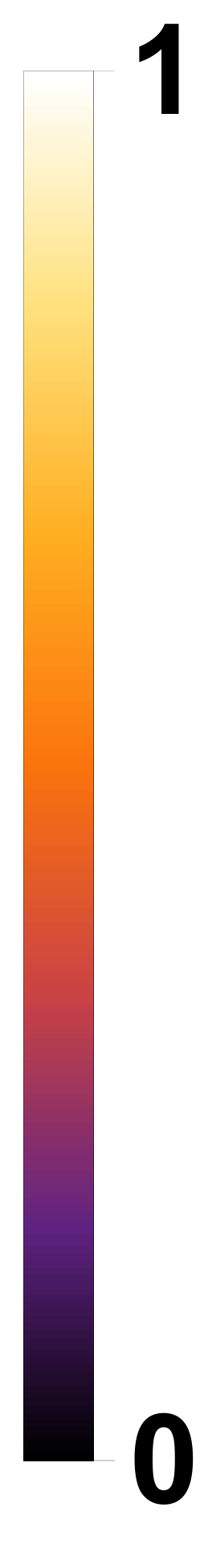}\\
			\qquad \\ \qquad \\}
	\caption{IPR for different eigenstates (arranged by $ E_m<E_{m+1} $), averaged over phases $ \varphi_{1,2} $ for each eigenstate. To check possible finite-size effects, we compute two system sizes $ L=40 $ in (a) and $ L=80 $ in (b), for the same parameter $ \theta=\pi/5 $. The green dashed line is the dual point $ V_d^{(D)} = 4 $. \label{fig:ipreig}}
	\end{figure}

	The inverse participation ratio (IPR) is defined as
	\begin{align}
	\text{IPR}_m = \sum_{x,y=1}^L|\phi_{x,y}^{(m)}|^4,
	\end{align}
	for a normalized eigenstate $ m $: $ \sum_{x,y}|\phi_{x,y}^{(m)}|^2 = 1 $. The IPR crosses from 0 in the strongly delocalized (i.e. take $ |\phi_{\boldsymbol{r}}| = 1/L $) regime to 1 in the localized situation (i.e. take $ |\phi_{\boldsymbol{r}}| = \delta_{\boldsymbol{r},\boldsymbol{r}'} $). See results in Fig.~\ref{fig:ipreig}. As mentioned previously, we average over various $ \varphi_1, \varphi_2 $'s in Eq.~(\ref{newmoire}) so as to reduce finite size effects concerning microscopic details. 
	
	From the results, there does seem to be mobility edges as we see that the transition of IPR for different eigenstates appears at different critical $ V_d^{(c)} $'s. Since the localization lengths do not need to be uniform, it is worthwhile to check whether some eigenstates have much longer localization lengths than others such that they appear like an extended one for a finite-size sample. For this purpose, we compute systems of different sizes for the same parameter in Fig.~\ref{fig:ipreig}. Should such a scenario occur, those localized states with larger localization legnths would take larger IPR values as one increases the system size (brighter colors should occupy larger areas). But there appears no notable finite-size effects as Fig.~\ref{fig:ipreig} (a) and (b) for sizes $ L=40, 80 $ look almost identical. Thus, the IPR results indeed suggest the existence of mobility edges for the two-dimensional generalization of Aubry-Andre model.
	
	Further, we note that the IPR feature here is quite different from what is typically seen in its one-dimensional counterparts. Here, the transition of IPR for different eigenstates changes continuously, constrained only by the particle-hole symmetry (after $ \varphi_{1,2} $ averages, or in $ L\rightarrow\infty $ limits) around $ E_{m_0}=0 $ for the middle state $ m_0=L^2/2 $. That is, there appear to be no ``plateaus'' of constant $ V_d^{(c)} $ for certain energy windows. This is to be compared with the one-dimensional scenario where only a few $ V_d^{(c)} $ exits, related by a ``step''-like transition at certain $ E_m $'s (see, e.g. Ref~\cite{Li2017}). 
	
	We trace such a difference back to the lack of gaps in the density of states for our two-dimensional models. In the one-dimensional system with mobility edge, large gaps (comparable with band width) typically exist in the density of states. Usually, an abrupt change of $ V_d^{(c)} $ occurs when the eigenstates go from one gapped band to another, while $ V_d^{(c)} $ remains almost constant within one band~\cite{Li2017}. In our two-dimensional case, we note that for small size systems, such as the $ \theta=\pi/5 $ example above for $ L\le 20 $, gaps in the density of states do appear, together with ``plateaus'' of $ V_d^{(c)} $ for $ E_m $ within one band. However, as the system size $ L $ increases, those gaps in the density of states shrink, and eventually vanish together with the ``plateaus" of $ V_d^{(c)} $. Then, a smooth change of IPR affected by density of states at $ E_m $ takes over, with typical features shown in Fig.~\ref{fig:ipreig}. After this point, the IPR configurations no longer change notably with the increase of $ L $. Such a scenario shows up for all cases we have tried, including various rotation angles $ \theta $ and stretched lattice constants for perturbing potentials. Thus, we expect that for a generic two-dimensional, quasi-periodic system in the thermodynamic limit, no gapped structure with ``plateaus'' of $ V_d^{(c)} $ should appear. The lack of gaps in the density of states is also shown for three-dimensional quasi-periodic models in Ref.~\cite{Devakul2017}. It is interesting to note that such a ``mini-band'' type of finite-size effects also exists in certain strongly interacting systems~\cite{Papic2015}. As such, for all results discussed in this work, we have made sure that the system size is well above the limit for band gaps to appear.

	\subsection{Critical Exponent}

\begin{figure}
		[h]
		\includegraphics[width=8cm]{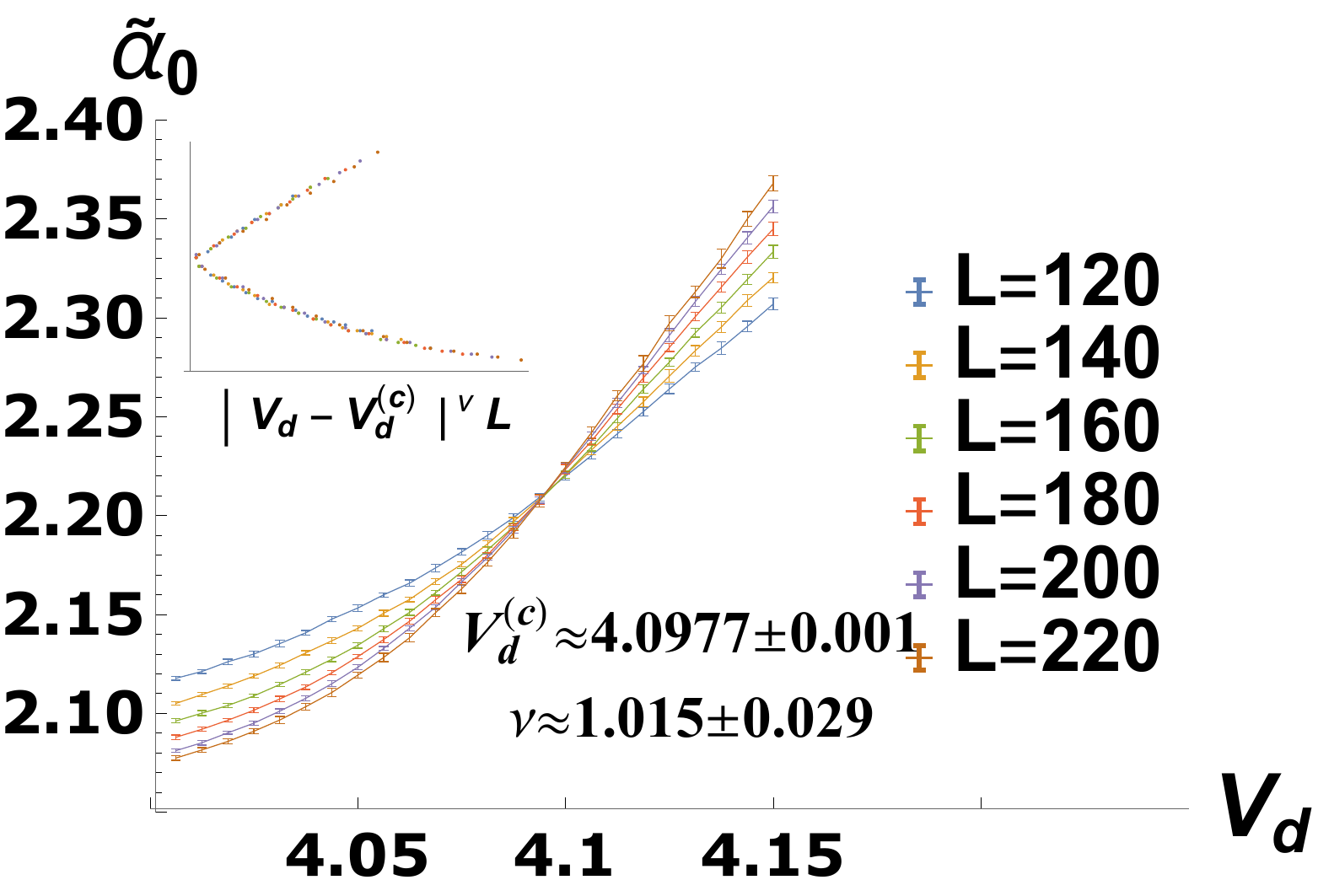}
		\caption{\label{fig:boxscaling} The scaling of $ \tilde{\alpha}_0 $ at energy $ E=0 $. The inset is the data collapse of $ \tilde{\alpha}_0 $ as a function of $ L/\xi \propto |V_d-V_d^{(c)}|^\nu L $ for various system sizes $ L $. The number of $ \varphi_1, \varphi_2 $ averaged over are $ 4\times 10^4 $ for $ L=120, 140, 160, 180 $ and $ 2\times 10^4 $ for $ L=200,220 $ respectively. The stable fit is given by expansion of Eq.~(\ref{eq:alpha1exp}) to the 5-th order. There are totally $ N_D=144 $ data points, $ N_p=8 $ fitting parameters, and the goodness of fit is evaluated by $ \chi^2=134  $, with $ p $-value $ 0.52 $. (See Appendix~\ref{app:fit} for more fitting details.)  }
	\end{figure}

	Near the critical point $ V_d^{(c)} $, the localization length is expected to diverge
	\begin{align}
	\xi \sim |V_d - V_d^{(c)}|^{-\nu}
	\end{align}
	with the scaling exponent $ \nu $. For the two-dimensional quasi-periodic model in the orthogonal class, such an exponent has been evasive in the previous studies~\cite{Devakul2017,Sutradhar2018}. Here, by sampling a large number of systems with different $ \varphi_{1,2} $'s based on the ``multifractal'' analysis, we obtain such an exponent as shown in Fig.~\ref{fig:boxscaling}. One advancement of our work is that the sizes $ L $ considered are systematically larger than those in previous works. For instance, the system sizes in Ref.~\cite{Devakul2017} ranges over $ L=40\sim120 $, while Ref.~\cite{Sutradhar2018} uses $ L=10\sim 90 $ in the exact diagonalizations for two-dimensional systems. We noticed that in our example of $ \theta=\pi/5 $, for size $ L\sim 20, 40, 60 $, there are severe finite-size effects as $ \tilde{\alpha}_0 $ defined below in Eq.~(\ref{eq:alphat}) shows irregular crossovers for different system sizes. As we tested, such finite size effects are much more severe than the one-dimensional Aubry-Andre model or a factorizable two-dimensional one ($ \theta=0 $ but with incommensurate stretching) where systems with similar sizes $ L $ already provide good scaling behaviors. In the following, let us describe the procedure of extracting $ \nu $ in more detail.
	
	The method of multifractal finite size scaling (MFSS) was introduced in Refs.~\cite{Rodriguez2009, Rodriguez2010, Rodriguez2011}. The steps can be summarized as follows. (1) Partition the system of size $ L\times L $ into boxes of size $ l\times l $, each containing $ (L/l)^2 $ lattice sites. In our case, $ L/l=10 $, so there are totally 100 boxes. (2) Obtain the real-space eigenstate $ \phi_{x,y} $ with energy close to certain value, i.e. $ E=0 $ in our case. (3) Compute the box-averaged value $ \tilde{\alpha}_0 $, where we have avoided boxes on the boundary so as to reduce finite-size effects,
	\begin{align}\nonumber
	\tilde{\alpha}_0 &= \frac{1}{((L/l)-2)^2} \frac{\sum_{a,b=2}^{(L/l)-1} \ln A_{a,b}}{\ln(l/L)}, \\ \label{eq:alphat}
	A_{a,b} &= \sum_{x=(a-1)l+1}^{al} \sum_{y=(b-1)l+1}^{bl} |\phi_{x,y}|^2.
	\end{align}
	Here, $ A_{a,b} $ is the wave function amplitudes within the box indexed by $ (a,b) $. For each system size $ L $, we need to diagonalize a large number of Hamiltonians with different $ \varphi_1, \varphi_2 $. Each of such a diagonalization (using Lanczos method) would render one eigenstate $ \phi_{x,y} $ closest to $ E=0 $ and its corresponding $ \tilde{\alpha}_0 $. Those $ \tilde{\alpha}_0 $'s for the same system sizes and $ V_d $'s are to be averaged over to provide one data point in Fig.~\ref{fig:boxscaling}. As emphasized in Ref.~\cite{Rodriguez2011}, only one sample state $ \phi_{x,y} $ should be used in each diagonalization.

	The qualitative behavior of $ \tilde{\alpha}_0 $ can be expected as follows. When the system is completely delocalized, i.e. wave function amplitudes are homogeneously distributed among all sites/boxes, $ |\phi_{x,y}|^2\sim 1/L^d $, $ A_{a,b}\sim (l/L)^d $, so $ \tilde{\alpha}_0\rightarrow d = 2 $ where $ d $ is the spatial dimension. In contrast, when approaching strongly localized regime with most of the boxes empty $ A_{a,b}\rightarrow0 $, $ \tilde{\alpha}_0\rightarrow \infty $. Thus, near the critical regime when the system crosses from metallic to insulating behavior at certain energy $ E $, $ \tilde{\alpha}_0 $ is a monotonically increasing function of $ V_d $. 
	
	The single-parameter scaling form can be written as
	\begin{align}\label{eq:alpha1exp}
	\tilde{\alpha}_0 = f(L/\xi) = g\left((V_d-V_d^{(c)})L^{1/\nu}\right),
	\end{align}
	which suggests the data collapse of $ \tilde{\alpha}_0 $ in terms of $ L/\xi\propto |V_d-V_d^{(c)}|^\nu L $. Near the critical point, we expand $ \tilde{\alpha}_0\approx \sum_{n=1}^N a_n w^n $, where $ w = (V_d - V_d^{(c)}) L^{1/\nu} $, and fit the data for different $ V_d $ and $ L $ with the polynomial. The quality of fit is evaluated by the $ \chi^2 $ statistics and the $ p $-value, and the uncertainty range is generated by a Monte-Carlo type of procedure using synthetic data sets having the same mean and standard deviation as the original data. Detailed fitting procedures are described in Appendix \ref{app:fit}. 
	
	The results are shown in Fig.~\ref{fig:boxscaling}, where we see that the scaling exponent $ \nu \approx 1 $ appears to saturate the Harris bound $ \nu>2/d=1 $ in two dimensions. To verify the generality of the scaling exponents, we computed another twist angle $ \theta=4\pi/9 $ in Appendix~\ref{app:pi94}, where $ \nu\approx1 $ is similarly obtained. It is also of interest to see that despite the existence of mobility edge in the two dimensional model, its critical exponent $ \nu $ is similar to the one-dimensional Aubry-Andre one with analytically obtainable $ \nu=1 $, where mobility edge is absent. As such, we can summarize the critical behavior for the self-dual models in various dimensions in Table~\ref{tab:critical}. 
	\begin{table}[h]
		\caption{\label{tab:critical} The critical behavior for self-dual, orthogonal class of models in a quasi-periodic potential.}
	\begin{tabular}{|l|c|c|c|}
		\hline 
		Dimension & Mobility edge? & Exponent $\nu$ & Harris bound satisfied? \\ 
		\hline 
		1~\cite{Aubry1980} & No & 1 & No \\ 
		\hline 
		2 & Yes & $ \approx 1 $ & Yes (saturate) \\ 
		\hline 
		3~\cite{Devakul2017} & Yes & $ \approx 1.6 $ & Yes (well above) \\ 
		\hline 
	\end{tabular} 
	\end{table}

	\section{Experimental Signatures}
	
	In principle, one could surely produce exactly the model in Eq.~(\ref{moire}) by superposing a standard square optical lattice together with the Moire perturbing potential given by a digital mirror device, like in the experiment Ref.~\cite{Choi2016}. Even without using the digital mirror device, a standard superlattice technique could realize a generalized version of the previous model. For the generality and broadness of our results, we would focus on such a generalized model in the following.

	\begin{figure}
		[h]
		\parbox{5.5cm}{\includegraphics[width=6cm]{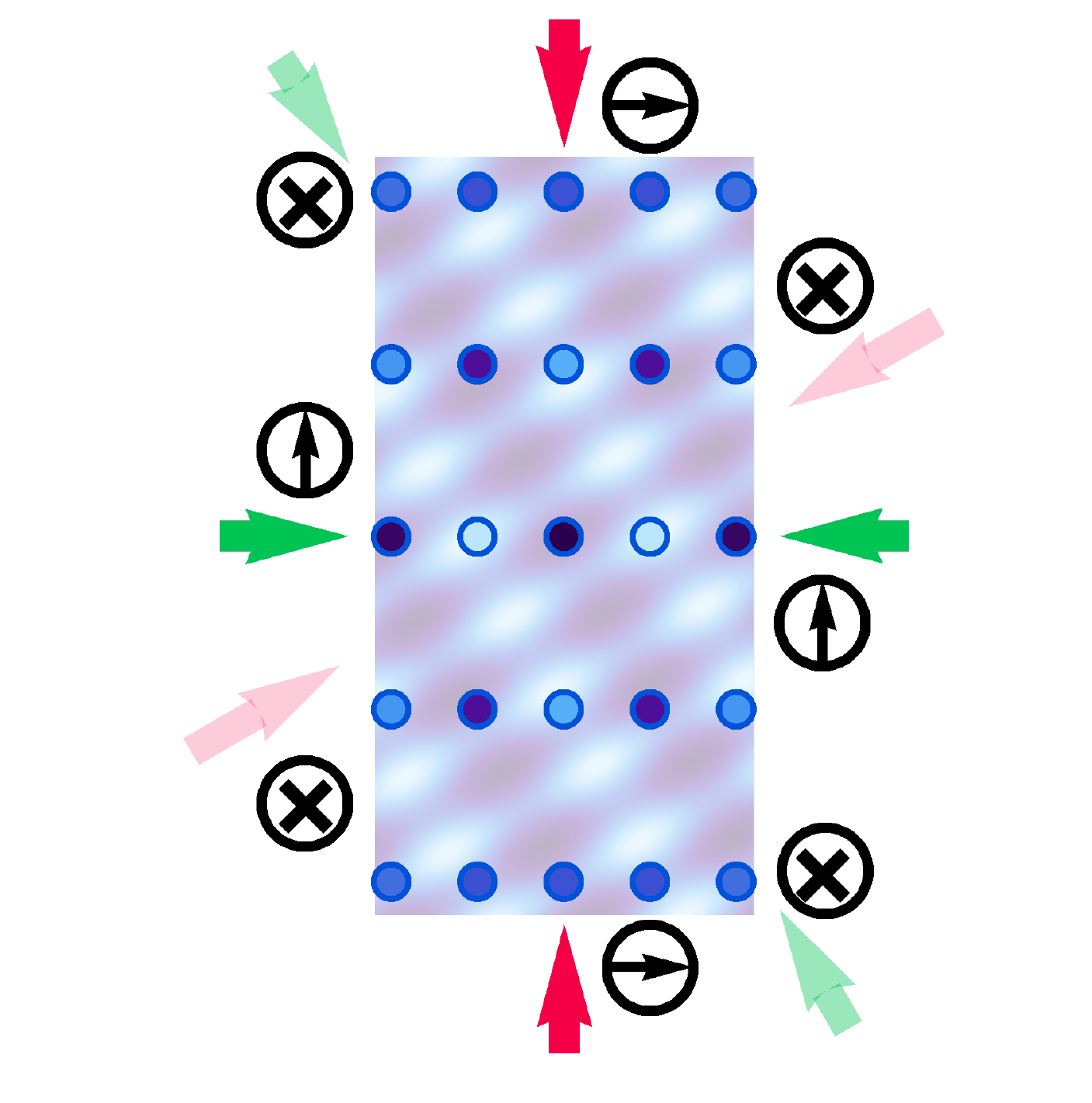}}
		\parbox{2.5cm}{\includegraphics[width=2.5cm]{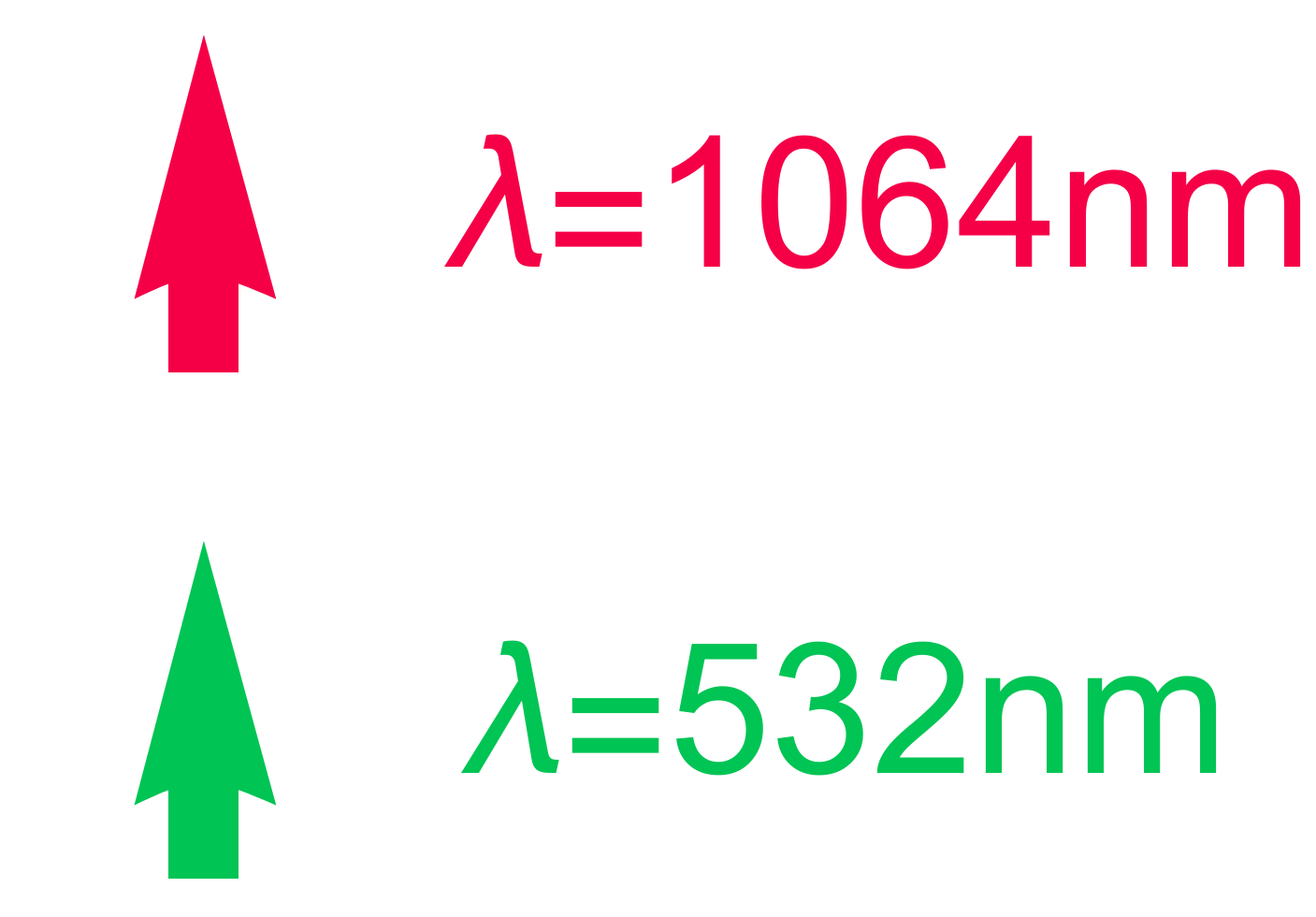}}
		\caption{\label{fig:expt} The experimental scheme for the Moire lattice. Black arrows and crosses denote the polarization directions such that the four pairs of laser beams do not interfere with each other. Here $ \theta=\pi/3 $. The random phases are $ \varphi_1=3, \varphi_2=2 $ chosen as an example.}
	\end{figure}
	In the superlattice scheme, we need two identical and mutually independent lattice potentials (without interference) overlapping with each other. Note that two square lattices involve 4 pairs of laser beams of the same wavelengths. Since there are only 3 perpendicular directions for polarizations, at least two pairs of the beams would end up interfering with each other and therefore the two sets of square lattice will not be mutually independent. To circumvent this difficulty, we consider a generalization to two identical {\em rectangular} lattices, where laser beams of very different frequencies will not interfere with each other when their effects are averaged over time. That means the Moire potential becomes 
	\begin{align}\nonumber
	\tilde{\mu}_{x,y} &= \sin^2 [\pi \tilde{u}_{x,y} - \varphi_1] + \sin^2[(\pi/\gamma)\tilde{v}_{x,y} - \varphi_2],\\
	\tilde{u}_{x,y} &= x\cos\theta - y \gamma \sin\theta, \qquad
	\tilde{v}_{x,y} = x\sin\theta + y \gamma\cos\theta,
	\end{align}
	with $ x,y\in\mathbb{Z} $ still denoting site indices. The self-duality mapping can be similarly performed by replacing $ u_{x,y}, v_{x,y} \rightarrow \tilde{u}_{x,y}, \tilde{v}_{x,y}/\gamma $ in Eq.~(\ref{eq:fourier}), and therefore the self-dual point is still $ V_d^{(D)}=4 $.
	In the following, we choose the aspect ratio of $ \gamma=2 $ for a plaquette of the rectangular lattices. This is to emphasize that there is no incommensurability arising from the lattice constants of two overlapping lattices, and also, such an aspect ratio can be achieved by using lasers with wavelengths $ \lambda_1 = 1064nm  $ and $ \lambda_2 = 532nm $ as in many experiments. A schematic plot for the lasers and their polarization directions are shown in Fig.~\ref{fig:expt}. Note that the commensurate condition Eq.~(\ref{commCond}) still holds for rectangular lattices with rational aspect ratios, as discussed in Appendix~\ref{app:comm}.

	\begin{figure}
		\includegraphics[width=8cm]{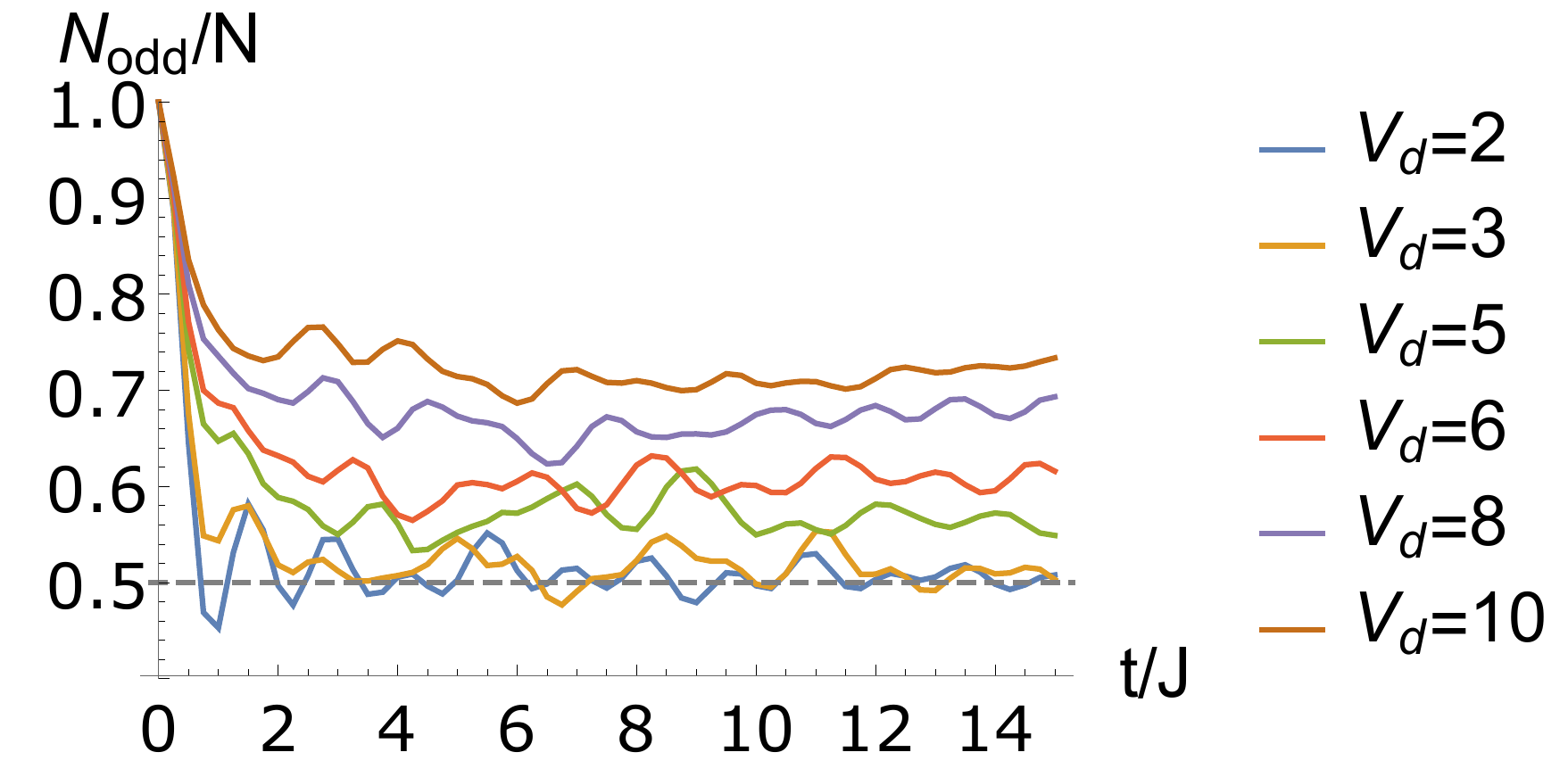}
		\caption{\label{fig:expt_cdw} The portion of particles remaining in odd $ x=1,3,5\dots $. Here $ \theta=\pi/3 $, and $ L=80 $.}
	\end{figure}
	Consider a spin-polarized fermion system. The initial state can be chosen similar to the previous experiments as a charge-density-wave type~\cite{Schreiber2015}, where even sites $ x=0,2,4,\dots $ along a direction are empty, while odd sites $ x=1,3,5,\dots $ are occupied. Fig.~\ref{fig:expt_cdw} shows the numerical simulation for the time evolution, where $ N_{\text{odd}} $ are particles at sites $ x=1,3,5\dots $ for arbitrary $ y $, and $ N $ is the total particle number. We readily see that close to self-duality point $ V_d^{(D)}=4 $, i.e. $ V_d=3 $, there already appear a significant portion of localized states such that $ N_{\text{odd}}/N $ remain slightly above $ 0.5 $, which signals the mobility edge in such a system.

	\section{Conclusion}
	
	We have provided, as the first step, a study on effective models of Moire localization in two-dimensional quasi-periodic systems. The commensuration conditions discussed in Sec.~\ref{sec:comm} can be used as a guide to engineer bilayer or multi-layer systems made of simple rectangular/square lattices, which is of primary interest to cold atom experiments. 
	
	Such a model also fills up the gap for the understanding of two-dimensional quasi-periodic system in the orthogonal class (time-reversal invariant). We find that the mobility edge here, shown by the IPR, appears quite differently from that in a one-dimensional quasi-periodic system studied before. In our case, there exists a rapid and continuous change without ``plateaus'' for IPR in the eigenstate-disorder plane. We trace such a difference to the lack of band gaps for the two-dimensional spectrum. Note that such a feature is most notably revealed in the IPR calculation rather than the level statistics, as the latter one would inevitably involve an average over many nearby eigenstates. On the other hand, level statistics reveals a cross-point for relatively small system sizes but appear to converge for larger $ L $'s, with the region well above duality point $ V_d=4 $ still far away from the Poisson limit. Such a fact also confirms the existence of mobility edge.

	
	We have also found the scaling exponent for such a model, which has been absent in previous studies~\cite{Devakul2017}. It also resolves the earlier concern in Ref.~\cite{Sutradhar2018} that a single-parameter scaling may be insufficient for this class of model, as we found that choosing sufficiently large system sizes would yield satisfactory scaling behaviors. The exponent for our two-dimensional model, $ \nu\approx 1 $, turns out to be the same as its one dimensional cousin, while the difference is that in two-dimensions such an exponent saturates the Harris bound $ \nu>2/d=1 $ formulated for a purely random scenario. Also, the exponent in our orthogonal class is smaller than the two-dimensional systems in symplectic classes $ \nu\approx 1.38 $~\cite{Devakul2017}.
	
	Finally, we have provided a scheme for cold atom experiments relying solely on standard superlattice techniques, without using the digital mirror device.  Since the system is only sensitive to the relative rotation angles rather than lattice constants between main and perturbing potentials, our results indicate that the two dimensional system is actually more viable for cold atom experiments as there are no requirements for particular laser wavelengths.

	\section{Acknowledgment}
	This work is supported by AFOSR Grant No.  FA9550-16-1-0006, MURI-ARO Grant No. W911NF-17-1-0323, and NSF of China Overseas Scholar Collaborative Program Grant No. 11429402 sponsored by Peking University. This work used the Bridges system, which is supported by NSF award number PHY190004P at the Pittsburgh Supercomputing Center (PSC).

	\appendix
	
	\section{Commensurate rotations for rectangular Lattices}\label{app:comm}
	Take an arbitrary lattice site as the center of rotation, then the commensurate rotations should satisfy
	\begin{align}
	m_1\hat{x} + \gamma m_2 \hat{y} = n_1 \hat{x}'+ \gamma n_2\hat{y}',
	\end{align}
	where $ \hat{x}' = \hat{x}\cos\theta  + \hat{y}\sin\theta  $, $ \hat{y}' = -\hat{x}\sin\theta  + \hat{y}\cos\theta  $, and $ \gamma $ is the aspect ratio for the rectangular lattices. The following derivations are valid for a {\em rational} $ \gamma $, i.e. $ \gamma=2 $ for the experiment part in the main text. Written explicitly,
	\begin{align}\label{rotationA}
	\begin{pmatrix}
	m_1 \\ \gamma m_2
	\end{pmatrix} = 
	\begin{pmatrix}
	\cos\theta & -\sin\theta \\
	\sin\theta & \cos\theta
	\end{pmatrix}
	\begin{pmatrix}
	n_1 \\ \gamma n_2
	\end{pmatrix}.
	\end{align}
	To have integer solutions $ (m_1,m_2), (n_1, n_2) $, it is necessary to have 
	\begin{align}\label{costheta}
	&&\cos\theta = k_1/k_3,&&
	\sin\theta = k_2/k_3, && k_1,k_2,k_3\in \mathbb{Z}.
	\end{align}
	The integers $ k_1,k_2,k_3 $ satisfy 
	\begin{align}\label{tripleEq}
	k_1^2 + k_2^2 = k_3^2.
	\end{align}
	The above diophantine equation corresponds to finding Pontryagin triples whose solutions are well-known~\cite{diophantinebook,Steuding2005,Shallcross2010}. One can enumerate them by the following procedure. Note that Eqs.~ (\ref{costheta}) and (\ref{tripleEq}) mean finding the rational points on the unit circle $ \cos^2\theta + \sin^2\theta = 1 $. One such point can be pinned down as $ (0,1) $, and the remaining ones can be found by intersecting the circle with the line passing through rational points $ (0,1) $ and $ (q/p,0) $, where $ q,p\in\mathbb{Z} $. 
	Combining (\ref{tripleEq}) and $ k_1/k_3 = 1- (q/p) k_2/k_3 $, we can set
	\begin{align}\label{kpq}
	k_1 = q^2- p^2, && k_2 = 2pq, && k_3 = q^2+p^2, && p,q\in\mathbb{Z}.
	\end{align}
	Then we have Eq.~(\ref{commCond}) in the main text. Note here $ k_1, k_2 $ are interchangeable, which correspond to angles $ \theta $ and $ \pi/2-\theta $.
	
	Note that the commensuration condition is the same for arbitrary aspect ratios $ \gamma $. But the Bravais lattice vectors are different, as we discuss below.

	\section{Moire Bravais lattice vectors at commensurate rotation angles}\label{app:vec}
	Define the matrices
	\begin{align}\nonumber
	& S=\frac{1}{\sqrt{q^2+p^2}}\begin{pmatrix}
	q & p\\ -p & q
	\end{pmatrix} = \frac{q}{\sqrt{q^2+p^2}}\mathbb{I} + \frac{p}{\sqrt{q^2+p^2}} i\sigma_y, \\
	& S^{-1} = S^T, \\
	&
	A = \frac{1}{q^2+p^2}\begin{pmatrix}
	q^2-p^2 & -2qp \\ 2qp & q^2-p^2
	\end{pmatrix} = \frac{q^2-p^2}{q^2+p^2}\mathbb{I} - \frac{2qp}{q^2+p^2} i\sigma_y
	\end{align}
	where $ ^T $ is transpose, and $ A $ is the rotation matrix in Eq.~(\ref{rotationA}) parametrized by Eq.~(\ref{costheta}) and (\ref{kpq}). They satisfy the relation $ SAS = \mathbb{I} $. Then the vectors $ \boldsymbol{m}, \boldsymbol{n} $ can be written as
	\begin{align}
	\begin{pmatrix}\label{eq:bv1}
	n_1\\\gamma n_2
	\end{pmatrix} &=\alpha \begin{pmatrix}
	q\\ -p 
	\end{pmatrix} + \beta \begin{pmatrix}
	p \\ q
	\end{pmatrix},\\ \label{eq:bv2}
	\begin{pmatrix}
	m_1 \\ \gamma m_2 
	\end{pmatrix} &= \alpha \begin{pmatrix}
	q \\ p
	\end{pmatrix} + \beta \begin{pmatrix}
	-p \\ q
	\end{pmatrix}
	\end{align}
	where $ \alpha,\beta $ are some rational numbers. Thus, the Moire Bravais vectors for a square lattice $ \gamma=1 $ can be chosen as
	\begin{align}
	\boldsymbol{T}_1 = \begin{pmatrix}
	q \\ -p
	\end{pmatrix}, &&
	\boldsymbol{T}_2 = \begin{pmatrix}
	p \\ q
	\end{pmatrix}
	\end{align}
	We aim to find the shortest vectors $ \boldsymbol{T}_{1,2} $ as it is possible that some linear combinations could yield
	\begin{align}
	\alpha_i \boldsymbol{T}_1 + \beta_i \boldsymbol{T}_2 = N_i \boldsymbol{T}'_i
	\end{align}
	where $ N_i\in \mathbb{Z} $ and $ \boldsymbol{T}'_i $ is also a Moire Bravais vector. Note here $ (\alpha_i,\beta_i) $ are coprime numbers, for otherwise the common factors can be canceled by $ N_i $.
	It turns out that the only exception is when $ (q,p) $ are both odd numbers, in which case 
	\begin{align}
	\boldsymbol{T}_1' = \frac{1}{2}\begin{pmatrix}
	q+p \\ q-p
	\end{pmatrix}, &&
	\boldsymbol{T}_2' = \frac{1}{2}\begin{pmatrix}
	q-p \\ -q-p
	\end{pmatrix}
	\end{align}
	
	For a rectangular lattice with rational aspect ratio $ \gamma $, one should apply the general conditions (\ref{eq:bv1}) to find the linear independent vectors $ \boldsymbol{T}=(n_1, n_2) $ such that $ \alpha q + \beta p $ and $ (-\alpha p + \beta q)/\gamma $ are integers, and $ q, p  $ are coprime. For our purpose, we give the lattice vectors for $ \gamma=2 $: when $ q, p $ are both odd, $ \boldsymbol{T}_1 = (q+p, (q-p)/2)^T, \boldsymbol{T}_2 = (-(q-p), (q+p)/2)^T $; when $ q $ is odd and $ p $ is even, $ \boldsymbol{T}_1 = (q,-p/2)^T, \boldsymbol{T}_2 = (2p, q)^T $; finally, when $ q $ is even and $ p $ is odd, $ \boldsymbol{T}_1 = (2q,-p)^T, \boldsymbol{T}_2 = (p, q/2) $.

	\section{Numerical algorithm for free fermion dynamics}
	
	Consider 
	\begin{align}
	H = \Psi^\dagger {\cal H} \Psi, && 
	\Psi = (c_1,\dots,c_N)^T,
	\end{align}
	with $ c_i $ the fermion operator at site $ i $ and $ {\cal H} $ an $ N\times N $ matrix. For an observable that can also be expressed in the bilinear form (such as the density)
	\begin{align}
	A = \Psi^\dagger {\cal A} \Psi, && A(t) = e^{iHt}Ae^{-iHt},
	\end{align}
	note $ [AB,C] = A\{B,C\} - \{A,C\}B $,
	\begin{align}
	e^{iHt} \Psi_\mu e^{-iHt} = \sum_{n=0}^\infty \frac{(it)^n}{n!} [(\Psi_\alpha^\dagger {\cal H}_{\alpha\beta} \Psi_\beta)^{(n)}, \Psi_\mu]= (e^{-i{\cal H}t})_{\mu\beta} \Psi_\beta,
	\end{align}
	and therefore 
	\begin{align}
	A(t) = \Psi^\dagger \left( e^{i{\cal H}t} {\cal A} e^{-i{\cal H}t} \right) \Psi.
	\end{align}
	Now, if the Hamiltonian matrix can be diagonalized by the matrix $ {\cal T} $,
	\begin{align}
	{\cal T}^\dagger {\cal H} {\cal T} = \text{diag}\{\varepsilon_1,\dots, \varepsilon_N\},
	\end{align}
	we have
	\begin{align}
	A(t) = \sum_{\alpha,\beta=1}^N \Gamma^\dagger_\alpha {\cal T}_\alpha^\dagger {\cal A}_{\alpha\beta} {\cal T}_\beta \Gamma_\beta e^{i(\varepsilon_\alpha - \varepsilon_\beta)t}, && \Gamma = {\cal T}^\dagger \Psi.
	\end{align}
	
	For our purposes, consider the initial state where half of the system is filled, $ |\psi_{\text{ini}} \rangle = c_1^\dagger\dots c_{N/2}^\dagger |0\rangle $. Then
	\begin{align}
	\langle\psi_{\text{ini}} | A(t) |\psi_{\text{ini}} \rangle = \sum_{i=1}^N n_i^{(0)} {\cal T}_{i\alpha} e^{i\varepsilon_\alpha t} \left( {\cal T}^\dagger {\cal A} {\cal T} \right)_{\alpha\beta } e^{-i\varepsilon_\beta} {\cal T}^\dagger_{\beta i},
	\end{align}
	where $ \alpha,\beta $ are summed over all eigenstates, and $ n_i^{(0)} $ is the initial particle number at site $ i $. The above formula applies to both free bosons and fermions when the initial state is a Fock one, with the restrictions that for fermions, $ n_i^{(0)}\le1 $.

	\section{Review of Aubry-Andr\'{e} model in one-dimension}\label{app:AAmodel}
	
	To be self-contained and to help trace the origin for mobility edge in two-dimensional models, we review some features of the one dimensional Aubry-Andre model. There are two necessary conditions for having uniform localization lengths: (1) A one-dimensional model with nearest-neighbor hopping; (2) The duality between localization/delocalization in real/momentum spaces. Generalizing to higher-dimensions clearly violate condition (1) and therefore do not exhibit unique localization length. The following derivation is based on the discussions in Refs.~\cite{Herbert1971,Thouless1972,Aubry1980,Goldsheid2005}. 
	
	The Aubry-Andre model is
	\begin{align}
	H &= -t\sum_{m=1}^{N}(a_{m+1}^\dagger a_m + a_{m}^\dagger a_{m+1}) + \sum_{m=1}^N \mu_m a_m^\dagger a_m,\\
	\mu_m &= \frac{V_d}{2} \cos (2\pi q m + \varphi) , \qquad V_d^{(D)} = 4.
	\end{align}
	Performing a similar analysis as in Sec.~\ref{sec:duality}, we could similarly obtain the duality mapping with critical disorder strength $ V_d^{(D)} $. However, due to the one-dimensional nearest neighbor feature, one could related the localization length with density of states through the Thouless formula as shown below.

	Consider the real space matrix for Green's function
	\begin{align}
	G_{mn}(E) = \left( \frac{1}{E-H} \right)_{mn} \equiv \langle 0|a_m (E-H)^{-1} a_n^\dagger |0\rangle,
	\end{align}
	where $ m,n = 1,\dots,N $ denote lattice sites, $ |0\rangle $ is the vacuum. For an open boundary system, the matrix 
	\begin{align}\label{eq:EH}
	E-H = 
	\begin{pmatrix}
	E-\mu_1 & -t & 0 & 0 & \dots 0 \\
	-t & E-\mu_2 & -t & 0 & \dots 0\\
	0 & -t & E-\mu_3 & -t & \dots 0\\
	\dots & \dots & \dots & \dots & \dots
	\end{pmatrix}.
	\end{align}
	{\em Due to the one-dimensional nearest-neighbor feature}, the corner element for the inverse matrix can be easily acquired as the cofactor
	\begin{align}\label{eq:G1N}
	G_{1N} = \frac{1}{\det (E-H)} (-t)^{N-1} (-1)^{N-1} = \frac{t^{N-1}}{\prod_{\alpha=1}^N(E-E_\alpha)},
	\end{align}
	where $ E_\alpha $ are eigenvalues of $ H $. On the other hand, according to the definition, one can first expand the Green's function in the energy eigenbasis $
	G(E) = (E-H)^{-1} = \sum_{\alpha=1}^N  (E-E_\alpha)^{-1} |\alpha\rangle \langle \alpha|$,
	and then the real space matrix element is
	\begin{align}
	G_{1N}(E) = \sum_{\alpha=1}^N \psi_{\alpha}(1) \psi_\alpha^*(N) (E-E_\alpha)^{-1},
	\end{align}
	where $ \psi_{\alpha}(m) = \langle 0|a_m|E_\alpha\rangle $ is the real-space eigenfunctions. Simultaneously using the above two equations to compute the residual of $ G_{1N}(E) $ at a certain $ E_\beta $,
	\begin{align}
	\lim_{E\rightarrow E_\beta} (E-E_\beta) G_{1N}(E) = \psi_\beta(1) \psi_\beta^*(N)  = \frac{t^{N-1}}{\prod_{\alpha, \alpha\ne\beta} (E_\beta- E_\alpha)}
	\end{align}
	Suppose the eigenstate $ |E_\beta\rangle $ is localized centering at some site $ m_0 $, with the ansats 
	\begin{align}
	\psi_\beta(1) = A_1 e^{-\lambda(E_\beta)(m_0-1)}, &&  \psi_\beta(N) =A_N e^{-\lambda(E_\beta)(N-m_0)}
	\end{align} 
	where $ \lambda $ is the exponent of interest, and $ A_m $ is the multiplication of some constant and non-decaying oscillating factors at site $ m $. Combing the above two equations and taking logarithm, we have
	\begin{align}
	\lambda(E_\beta) = \frac{1}{N-1}\left(
	\ln(A_1A_N) + \sum_{\alpha,\alpha\ne\beta} \ln|E_\beta-E_\alpha|
	\right) - \ln t.
	\end{align}
	Taking the $ N\rightarrow\infty $ limit, the $ \ln(A_1A_N) $ term vanishes. The second term can be converted into an integration. Then
	\begin{align}
	\lambda(E_\beta) = -\ln(t) + \int_{-\infty}^{\infty}  (\ln|E_\beta-\varepsilon|) d N(\varepsilon),
	\end{align}
	where $ d N(\varepsilon) = D(\varepsilon) d\varepsilon $ and $ D(\varepsilon) $ is the density of state at $ \varepsilon $. This is the Thouless formula for the localization length $ \xi = 1/\lambda $. Usually, the hopping constant $ t=1 $, so the first term drops out. This formula can be easily generalized to inhomogeneous nearest neighbor hopping situations~\cite{Thouless1972}. But note that once the hopping goes beyond nearest neighbor, i.e. for second-neighbor hopping or in higher dimensions, Eq.~(\ref{eq:EH}) no longer takes the same form, and Eq.~(\ref{eq:G1N}) cannot be obtained. Then the Thouless formula would not hold generally.
	
	Due to the duality mapping, one can obtain the relation 
	\begin{align}
	N_{t,V_d}(\varepsilon) = N_{t,\frac{8t^2}{V_d}} (\frac{4t\varepsilon}{V_d})
	\end{align}
	for an incommensurate system. Then, applying Thouless formula,
	\begin{align}
	\lambda_{t,V_d}(E_\beta) = \lambda_{t,\frac{8t^2}{V_d}}(\frac{4tE_\beta}{V_d}) + \ln\frac{V_d}{4t}.
	\end{align}
	Since $ \lambda\ge0 $ by definition, for $ V_d>4t $, $ \lambda_{t,V_d}(E_\beta) \ge \ln(V_d/4t)>0 $. That means all eigenstates $ |E_\beta\rangle $ are exponentially localized. From the transformation between eigenfunctions in two dual representations, all eigenfunctions in the dual space are then delocalized (localization/delocalization are opposite in the two spaces related by Fourier transformation), meaning $ \lambda_{t,8t^2/V_d}(4tE_\beta/V_d)=0 $. Then, we have the same critical exponent (inverse of localization length $ \xi $)
	\begin{align}
	\lambda_{t,V_d} = \frac{1}{\xi} =  \ln\frac{V_d}{4t}
	\end{align}
	for all eigenstates when $ V_d/4t>1 $.

	\section{Additional details for scaling exponents}
	
	\subsection{Fitting procedure}\label{app:fit}
	The procedure for fitting the scaling of $ \tilde{\alpha}_0(V,L) $ consists of three steps.
	
\begin{figure}
		[h]
		\includegraphics[width=2.7cm]{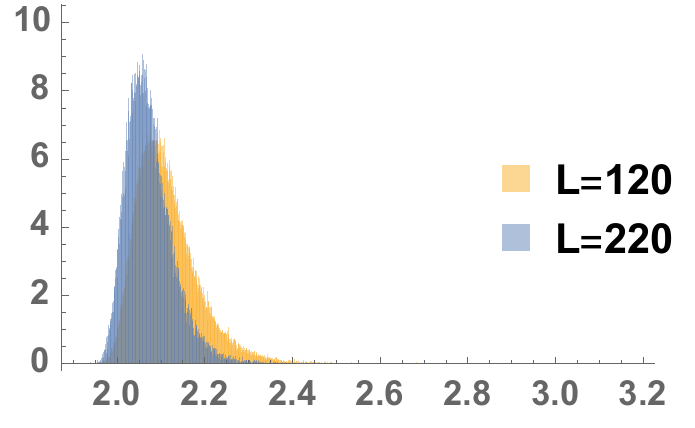}
		\includegraphics[width=2.7cm]{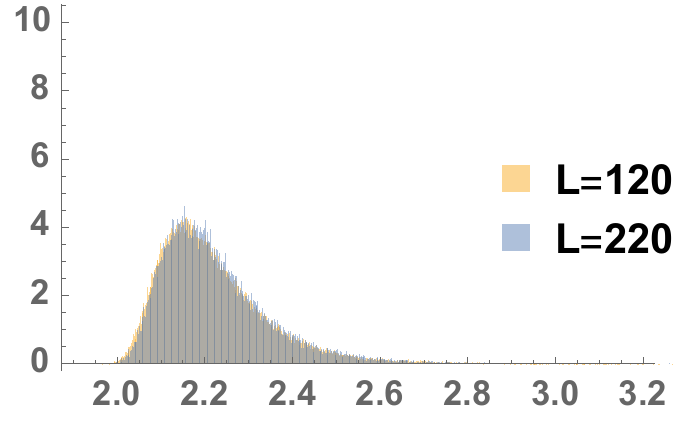}
		\includegraphics[width=2.7cm]{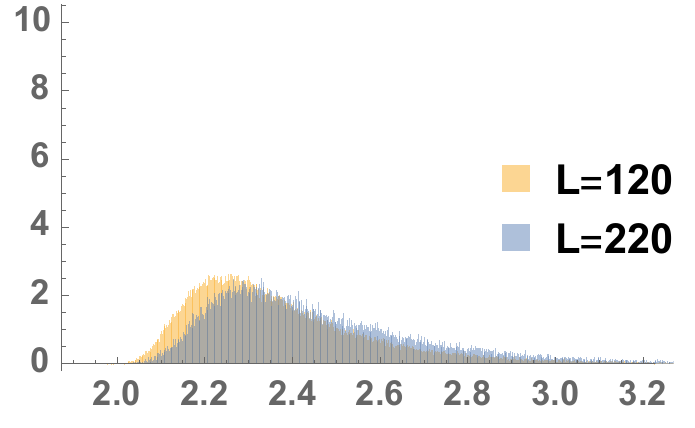}
		\caption{\label{fig:histogram_alpha} The (normalized) histogram for $ \tilde{\alpha}_0 $ at $ V_d=4.0 $ (left), $ V_d=4.1 $ (middle), and $ V_d=4.2 $ (right), computed using different $ \varphi_{1,2} $. For larger systems $ L $, the distribution inclines towards smaller ($ V_d $ in the metal limit) or larger ($ V_d $ in the insulator limit) values, while in the critical $ V_d $ they overlap.}
	\end{figure}

	First, the average $ \langle \tilde{\alpha}_0 \rangle (V,L) $ and standard deviation $ \sigma(V,L) $ of $ \tilde{\alpha}_0 $ at each point $ (V,L) $ is determined. There are  data of the order $ 10^5 $ at each $ (V,L) $ calculated by using different $ \varphi_1, \varphi_2 $. Note that although the average value is straightforward to understand, the ``standard deviation" must be taken with care. In all regimes, the $ \tilde{\alpha}_0 $ takes values according to a broad distribution function (see, i.e. Ref.~\cite{Rodriguez2011} Fig. 3). We reproduced such a character for our model in Fig.~\ref{fig:histogram_alpha}. It is in fact the maximal point of the distribution function that gives the scaling behavior. Only when averaged over large numbers of $ \varphi_{1,2} $, the mean value of $ \tilde{\alpha}_0 $ would approach the maximal probability point. Thus, at each $ (V,L) $, we divide all data into 10 bins, obtain the respective averaged $ \tilde{\alpha}_0 $, and compute the standard deviation using the 10 bin-averaged $ \tilde{\alpha}_0 $'s.
	
	Second, we fit $ \langle \tilde{\alpha}_0 \rangle (V,L) $ into the scaling function $ g(L/\xi) $ mentioned in the main text by minimizing $ \chi^2 $,
	$ \chi^2 = \sum_{V,L} (  \langle \tilde{\alpha}_0 \rangle(V,L) - g(V,L) )^2 / \sigma^2(V,L) $. For a reasonable fit, the $ \chi^2 $ value should be similar or less than the number of degrees of freedom, $ k = N_D - N_p $, where $ N_D $ is the number of $ \langle \tilde{\alpha}_0 \rangle (V,L) $ to be fit for different $ (V,L) $, and $ N_p $ is the number of fitting parameters. Whether or not a fit is acceptable is decided by the $ p $-value,
	\begin{align}
	p = \frac{\Gamma(k/2,\chi^2/2)}{\Gamma(k/2)},
	\end{align}
	where $ \Gamma(a,z) = \int_a^\infty t^{z-1}e^{-t}dt $ is the generalized Euler-Gamma function, with $ \Gamma(z) = \Gamma(0,z) $. A larger $ p $-value signals a better fit, and we take $ p\ge 0.1 $ as the acceptable criterion. In addition to the $ p $-value, we also make sure the fitting stability by checking that increasing the expansion order would result in $ V_d^{(c)}, \nu $ within the uncertainty range (to be discussed below). Under such criterions, the expansion order is kept as low as possible.
	
\begin{figure}
		[h]
		\includegraphics[width=4cm]{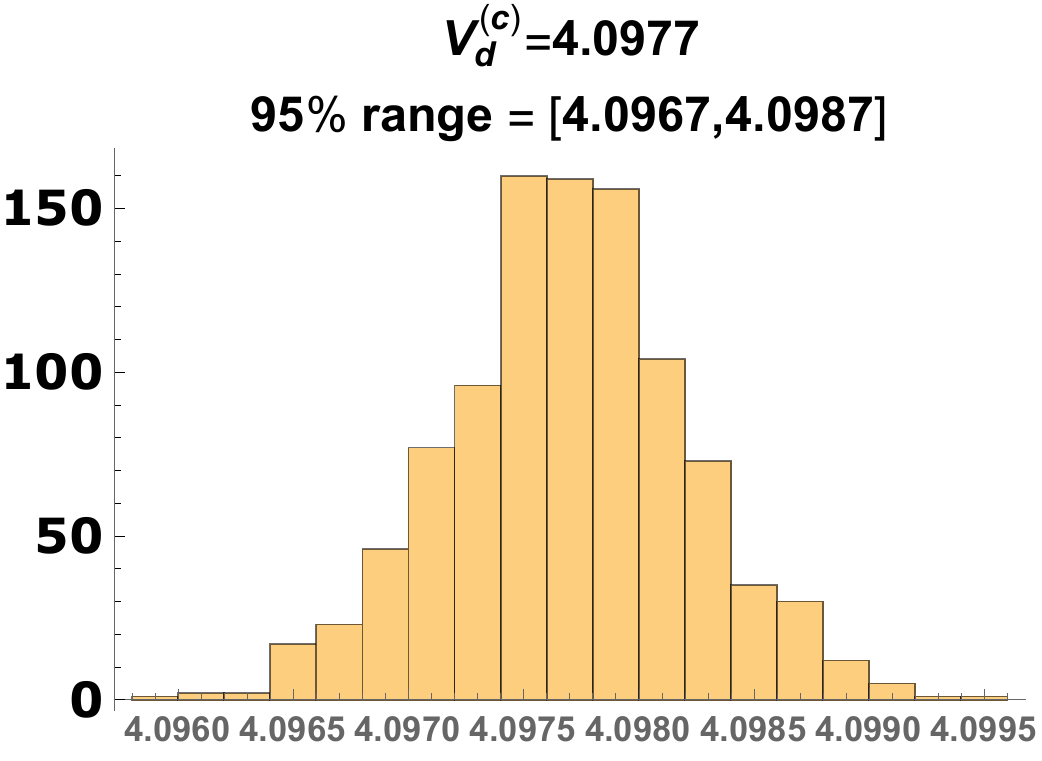}
		\includegraphics[width=4cm]{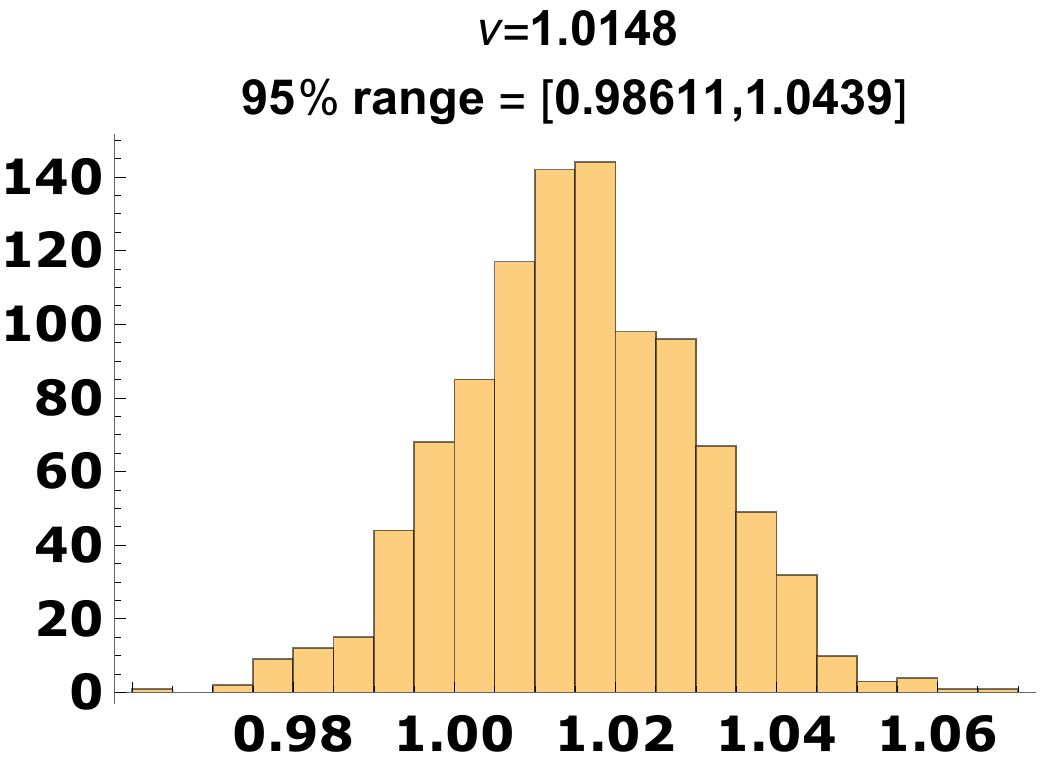}
		\caption{\label{fig:histogramvnu} The histogram for fitted $ V_d^{(c)} $ and $ \nu $ at the angle $ \theta=\pi/5 $, using 1000 synthetic data sets with the same standard deviations as in Fig.~\ref{fig:boxscaling} at each point. The $ 95\% $ confidence range is determined by discarding the $ 2.5\% $ of maximal/minimal data for $ V_d^{(c)}, \nu $ on the two sides.}
	\end{figure}

	Finally, once a best fit $ g(V,L) $ is obtained for $ \langle \tilde{\alpha}_0 \rangle (V,L) $ and $ \sigma(V,L) $, we generate 1000 synthetic data $ \alpha^{(syn)}_i(V,L), i=1,\dots,1000 $, at each point $ (V,L) $. They distribute according to the same mean and standard deviations discussed previously. Then, we fit each set of $ \{ \alpha^{(syn)}_i(V,L) | \forall V,L\} $ with the same expansion as in the best fit $ g(V,L) $, and obtain a histogram of $ V_d^{(c)}, \nu $ corresponding to $ i=1,\dots,1000 $. After discarding the ending (maximal and minimal) $ 2.5\% $ of data, the uncertainty range is obtained. The histogram of $ V_d^{(c)}, \nu $ for $ \theta=\pi/5 $ is shown in Fig.~\ref{fig:histogramvnu}.

	\subsection{Scaling at another angle }\label{app:pi94}
\begin{figure}
		[h]
		\includegraphics[width=7cm]{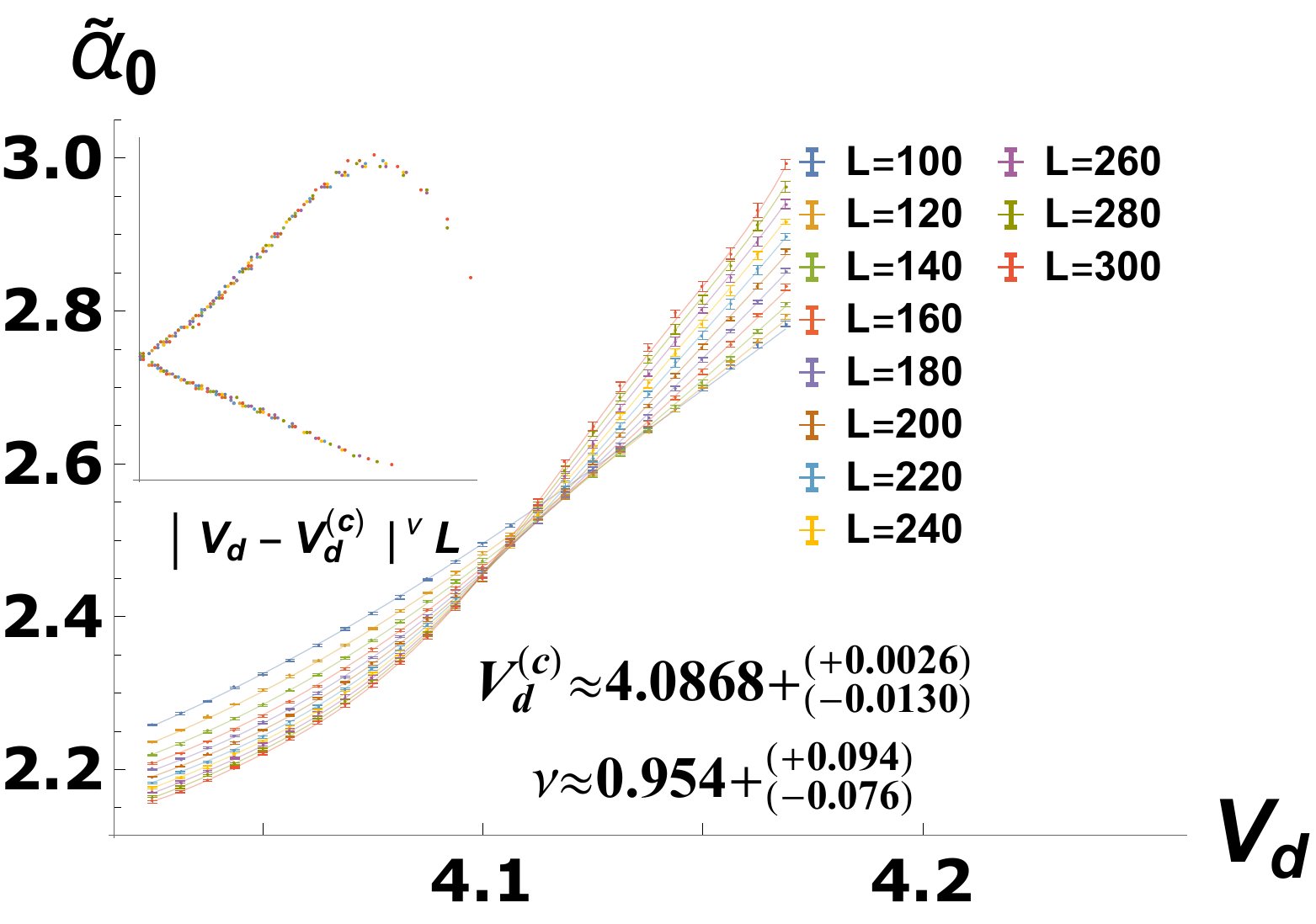}
		\caption{\label{fig:boxscaling_pi94} The scaling of $ \tilde{\alpha}_0 $ for $ \theta=4\pi/9 $ at $ E=0 $. The inset shows the scaling of $ \tilde{\alpha}_0 $ in terms of $ L/\xi\propto|V_d-V_d^{(c)}|^\nu L $ after subtracting the scaling-irrelevant term (the second term in Eq.~(\ref{eq:g0g1})). The numbers of $ \varphi_{1,2} $ being averaged over are $ 10^6 $ for $ L\le200 $ and $ 4\times10^5 $ for $ L\ge 220 $. The fitting data/parameters has $ N_D=262, N_p=16 $ (with $ N_1=N_2=6 $), and the quality of fit is evaluated by $ \chi^2=204, p=0.98 $. The uncertainty range is determined by synthetic data set as described previously. }
	\end{figure}

	We perform the same analysis for the angles $ \theta=4\pi/9 $ at $ E=0 $. But here we used a wider range of system lengths $ L $, and it is clear from Fig.~\ref{fig:boxscaling_pi94} that there is a shift in the ``crossing" point with the increase of $ L $. Such a ``shift" diminishes as $ L $ becomes larger, which means there is an additional scaling-irrelevant term entering $ \tilde{\alpha}_0 $~\cite{Rodriguez2011}. Therefore, we include such a term in the scaling function,
	\begin{align}
	\label{eq:g0g1}
	\tilde{\alpha}_0 = g_0\left((V_d-V_d^{(c)})L^{1/\nu} \right) + L^{-|y|} g_1\left((V_d-V_d^{(c)})L^{1/\nu} \right).
	\end{align}
	Compared with Eq.~(\ref{eq:alpha1exp}), the $ L^{-|y|} $ term describes the first order expansion of scaling-irrelevant term. With the fitting parameter $ -|y|<0 $, this term vanishes in the thermodynamic limit. Then we expand $ g_{0,1} $ similarly as before, $ g_0 = \sum_{m=0}^{N_0} a_mw^m $, $ g_1 = \sum_{m=0}^{N_1} b_mw^m $, where $ w=(V_d-V_d^{(c)}) L^{1/\nu} $. The fitting parameters are $ a_m, b_m, y, \nu, V_d^{(c)} $, with the total number $ N_p = N_0+N_1+5 $.
	
	\newpage

	%

\end{document}